\documentclass[11pt]{iopart}

\usepackage{iopams}
\pdfoutput=1
\usepackage[ansinew]{inputenc}
\usepackage{bm}
\usepackage[dvips]{graphicx}
\usepackage{color}

\newtheorem{propos}{Proposition}[section]

\newtheorem{coro}{Corollary}[section]

\begin{document}
\title[Causal characterization of singularities]{On the causal characterization of singularities in spherically symmetric spacetimes.}
\author{F. Fayos\thanks{Also at Laboratori de F\'{\i}sica Matem\`atica,
Societat Catalana de F\'{\i}sica, IEC, Barcelona.}  and R. Torres \\
Department of Applied Physics, UPC, Barcelona, Spain. \\
\eads{\mailto{f.fayos@upc.edu}, \mailto{ramon.torres-herrera@upc.edu}}}
\verb"Journal-ref: Class. Quantum Grav. 28 (2011) 215023"
\date{}

\begin{abstract}
The causal character of the zero-areal-radius ($R=0$) singularity in spherically symmetric spacetimes is studied. By using the techniques of the qualitative behaviour of dynamic systems, we are able to present the most comprehensive scheme so far to try to find out their causal characterization taking into account, and analyzing, the possible limitations of the approach. We show that, with this approach, the knowledge of the scalar invariant $m\equiv R(1-g^{\mu\nu}\partial_\mu R\partial_\nu R)/2$ suffices to characterize the singularity. We apply our results to the study of the outcome of Black Hole evaporation and show different possibilities. In this way, we find that a \emph{persistent} naked singularity could develop in the final stages of the evaporation and we show its distinctive features. Likewise, we study the options for the generation of naked singularities in the collapse of an object (such as a star) as a means of violating the cosmic censorship conjecture.
\end{abstract}


\section{Introduction}

Singularities are not part of the spacetime since they are related to diverging curvature invariants, to the incompleteness of curves and/or to the lack of tangent vectors.
At most, it would seem reasonable to say that singularities are situated in the \textit{boundary} of the spacetime, provided a suitable definition of boundary is given.
%
In fact, one such definition was first introduced by Penrose in 1963 \cite{Penrose63}. His idea was to embed the spacetime under study with metric $\mathbf g$ into another Lorentzian manifold (the \textit{unphysical} spacetime) with metric $\mathbf {\bar g} $ conformally,
\begin{equation*}
\mathbf g= \Omega^2 \mathbf{\bar g},
\end{equation*}
so that the causal properties are trivially kept.
In this way, the boundary acquires causal properties itself which are obtained by its mere examination in the unphysical spacetime. Specifically, it becomes now meaningful to give a singularity attributes such as spacelike, timelike or lightlike.
Furthermore, in spherically symmetric spacetimes,
where the $SO(3)$ group orbits form a spacelike two surface (the \textit{2-spheres}),
it is possible to perform just the conformal compactification of the two-dimensional surface orthogonal to the \textit{2-spheres} retaining all the important information. This is so because, by means of a coordinate change, the induced Lorentzian metric or \textit{first fundamental form} of the two-dimensional surface can always be brought into a conformally flat form
\begin{equation*}
ds^2_{2d}= \Omega^2 (x_0,x_1) (-dx_0^2+dx_1^2).
\end{equation*}
In this way, it can be naturally embedded in an {\it unphysical} two-dimensional Minkowskian spacetime (see, for instance, \cite{RepSeno}). This allows to draw simple two-dimensional diagrams, called Penrose diagrams, from which one can find out the properties of the boundary at a glance. In case the boundary is a $C^1$ curve at a given point in the Minkowskian spacetime one could causally characterize the point according to the tangent vector to the curve as usual. However, as the cases analyzed in the literature show\cite{exactET}, one usually finds boundaries that are just \emph{piecewise} $C^1$ and which provide us with ``piecewise causal characterizations".
%


In order to find out the main features of the local causal character of a singular boundary in a spherically symmetric spacetime it is not necessary to follow completely the conformal procedure explained above. This is interesting because an \emph{analytic} conformal compactification can only be found for certain particular cases. The alternative procedure is based on the general fact that the concept of a null geodesic is a conformally invariant one, so that for every null geodesic in the physical two dimensional surface there is a corresponding null geodesic in the unphysical two dimensional minkowskian spacetime\cite{BRP}. It seems reasonable, as we will try to show, that the study of the behaviour of the null geodesics in the physical spacetime in the neighborhood of a singularity will provide us with information about the behaviour of the geodesics around its corresponding conformal boundary and, as a consequence, on its causal character.

On the other hand, we will mainly deal with probably the most interesting type of singularities in spherically symmetric spacetimes: The \textit{zero-areal-radius}\footnote{Note that many authors call this singularity ``central", especially when studying the collapse of massive objects, regardless of its causal characterization. However, this terminology is hard to justify in general. For instance, the \textit{big-bang} singularity in a Robertson-Walker model is simply a particular case of a $R=0$ singularity, but it would hardly be called \textit{central}.} \textit{scalar curvature singularities} \cite{Seno}. In order to define the concept we will use the \textit{areal radius} $R$ such that the area of a 2-sphere is $4\pi R^2$.
Then, we say that there is a zero-areal-radius scalar curvature singularity at a point $p$ in $R=0$ if any scalar invariant polynomial in the Riemann tensor diverges when approaching it along any incomplete curve.

The difficulties to apply the above procedure will lie not in the identification of this type of singularities, but in the study of the behaviour of the radial null geodesics. In order to carry out this study we will analyze the system of differential equations that describe the null geodesics by means of
the standard \textit{qualitative theory of dynamic systems} \cite{Perko,Jordan,Nemy,Andronov,DLLA}. (Note also that we have written appendices A and B with the main results on this subject for the less usual cases). The information supplied by this theory about the behaviour of the radial null geodesics will allow us to mathematically classify the neighborhood of $p$ in the cases where the theory can be applied. As we will see, this together with the correct interpretation of the results for the $R\geq 0$ region, will provide us with a \textit{piecewise} characterization of the singularity around the chosen point.

But before putting this plan into practice, let us comment on the previous results and the relevance of this analysis. In fact, the study of $R=0$ singularities has been carried out for important particular solutions. In some cases the conformal boundaries have been obtained by using a conformal compactification (for example, Schwarzschild's and Reissner-Nordstr\"{o}m's solutions -see, for instance, \cite{H&E}- and Vaidya's solution when there is a linear mass function \cite{Volovich}\cite{HisWiEar}), while in other cases there is not analytical compactification and the alternative method of studying the radial null geodesics has been used in order to get the local causal characterization (for example, Vaidya's solution in the general imploding case \cite{Kuroda}\footnote{Incidentally, the reader can analyze this case, in which a singularity develops from a regular $R=0$ with $m=0$, to verify that the approach of studying the causal character of $\mathbf{n}=dR$ (or `` $g^{RR}$ ", provided the metric is given in suitable coordinates) in order to characterize a $R=0$-singularity on the boundary of the spacetime (!) is not a reliable method.} or so many different collapsing stellar models, like those found in \cite{E&S}\cite{Chris}\cite{O&P}).

In addition to the analysis of particular cases, some general approaches for studying zero-areal-radius singularities have also been carried out by analyzing the properties of the radial null geodesics. In particular, it has been shown \cite{Hayward} that
a $R=0$ singularity is spacelike (and trapped) at a point $p$ in $R=0$ if $m\rfloor_p>0$ and timelike (and untrapped) if $m\rfloor_p<0$, where $m\equiv R(1-g^{\mu\nu}\partial_\mu R\partial_\nu R)/2$.
Nevertheless, the case $m\rfloor_p = 0$, in which there could be either a regular center or a (spacelike, lightlike or timelike) singularity at $p$, must be analyzed in detail for every particular case. The trouble is that this is precisely the most interesting case in many different physical situations. For example, what is the outcome of the evaporation of a black hole if it gets rid of all of its mass? What transformations can the BH's singularity undergo in the process? And, since only particular cases have been treated in the current literature \cite{HisWiEar}\cite{HiscockEBH}\cite{Ring}\cite{F&NS}\cite{LocalBehav}, have all the possibilities been considered? On the other hand,
if a singularity-free star collapses, can it generate a massless singularity at the evaporation event due to the focusing of the different shells that constitute its interior? And, in the affirmative case, what can its causal character be?

In order to clarify the importance of these questions let us remind that
the singularity theorems \cite{Seno}\cite{H&E} show that, given a few reasonable assumptions, a collapse can terminate in a gravitational singularity. However, the theorems do not inform us about many properties of the singularities \cite{Seno}. Among others, there is lack of information about the type of singularity, the divergence of the energy density of matter fields and whether the singularity is hidden from outside view by the formation of a black hole. With regard to the last point, we must emphasize that the theorems do allow for the possibility that
the singularity could be seen by observers close enough to the singularity (in which case we say that there is a locally naked singularity) or from the future null infinity (globally naked singularity). In any case, there would be a hypersurface (the \textit{Cauchy Horizon}) beyond which General Relativity loses its predictability. The question on whether General Relativity contains a built-in safety feature that precludes the formation of naked singularities in generic gravitational collapses was put forward by Penrose in 1969 \cite{Penrose} and gave rise to what is known as the \textit{cosmic censorship conjecture} (CCC). Clearly, the causal character of the singularity is central in the resolution of this conjecture since, by construction, there are always null geodesics which are past incomplete whenever the spacetime possesses timelike or past null singularities (see, for example, section \ref{ngcc}). Therefore, timelike and past null singularities are always naked.

Some counterexamples to the cosmic censorship conjecture have been proposed.
An outstanding collapsing and radiating model can be found in the work by Demianski and Lasota \cite{D&L}. Even if it was not first proposed to be such a counterexample, but as an \textit{evaporating} model,
it was later shown \cite{SK&L} that it possesses an instantaneous naked scalar curvature singularity at the evaporating event.
Another counterexample of historical importance was discussed in \cite{Volovich}\cite{HisWiEar}\cite{Papapetrou} considering the collapse of null dust modelled by using Vaidya's solution.
It was shown that it suffices that pure radiation (or null dust) with a sufficiently weak wave travelling into an initially flat space-time focuses in $R=0$ in order to create a null singularity which is at least locally visible.
We will not intend now to exhaust all the different counterexamples to the cosmic censorship conjecture that can be found in the literature, but just to mention some of them we point out that naked singularities in spherically symmetric models are also possible in the collapse of dust \cite{E&S}\cite{Chris}\cite{Banerjee}\cite{Newm}, perfect fluids \cite{L&H}, general fluids \cite{D&J}, massless scalar fields \cite{Chris2} and even in higher dimensions \cite{B&C}.
On the other hand, more general studies on the formation of naked singularities in spherically symmetric spacetimes based on the study of the radial null geodesics can be found in \cite{Lake}, \cite{Singh} and, by using ad hoc devised procedure, in \cite{GGMP}.

Notwithstanding the above (incomplete) list of proposed counterexamples to the CCC, the subject is still open. This is so because, on the one hand, any specific example is unlikely to be considered generic in some appropriate sense and specific examples satisfying the CCC exist for the different matter fields above (see \cite{QCC}\cite{Joshicur} and references therein). Moreover, in \cite{Dafermos} the weak version of the CCC \cite{WCCC} has been shown to hold for a wide variety  of spherically symmetric coupled Einstein-matter systems. On the other hand, some examples possessing naked singularities have been shown to have some kind of instability. However, it is \emph{in general} unclear whether the instabilities can really hide the singularity (see \cite{QCC}\cite{Ori}\cite{Dotti}\cite{Nolan} and references therein).

Our aim in this paper is to apply the techniques of the qualitative behaviour of dynamic systems to the radial null geodesics around every $R=0$ singular point in order to study the causal characterization of the $R=0$ singularities. This will allow us to ascertain the relevant quantities (as well as their associated values) that determine whether a singularity is essentially spacelike, lightlike or timelike (even if the singularity is \textit{piecewise} timelike or \textit{piecewise} spacelike, for the case of piecewise $C^1$ boundaries). We will also find out and explicitly state the limits for the applicability of our results coming from our specific approach.
%
We would like to remark that ours is a geometrical approach requiring only the existence of a spacetime, but not the fulfillment of Einstein's equations.
Thus, we just try to discover the possibilities allowed by this geometrical approach which includes the classical as well as the semiclassical framework.
With the obtained information we will be able to study different possibilities for the final outcome of Black Hole evaporation and the different options for the generation of naked singularities. We will emphasize new models and possibilities that have not been taken into account so far.

The paper has been divided as follows: In section \ref{ngcc} we study the relationship between the null geodesics in the physical and the unphysical spacetimes and we revise how to extract information about the causal characterization from them. In section \ref{GSSM} we establish a general spherically symmetric spacetime and the conditions required for it to have a $R=0$ scalar curvature singularity as well as the equations governing its radial null geodesics. Section \ref{csmneq0} is an application for the trivial and well-known case of a non-zero function $m$ as $R$ tends to zero. In sections \ref{m0hyp} and \ref{m0nonhyp} we use the theory of qualitative behaviour of dynamic systems to deal with the analysis of isolated $m(R\rightarrow 0)=0$ \textit{points}, since, as we will see, they turn out to be the isolated critical points of the system of differential equations describing the radial null geodesics. Specifically, the hyperbolic and the non-hyperbolic cases are treated in sections \ref{m0hyp} and \ref{m0nonhyp}, respectively. In order to exhaust all the different possibilities, the non-isolated $m(R\rightarrow 0)=0$ \textit{points} are treated in section \ref{NICP}. The last section is devoted to the consequences and applications of our results. In particular, the general cases of Black Hole evaporation and the generation of naked singularities are analyzed.

\section{Null geodesics and causal characterization}\label{ngcc}


%

Let us assume that we are given an oriented two-dimensional Lorentzian manifold and that we embed it into an unphysical two-dimensional Minkowskian spacetime, as explained in the introduction. Provided that the singular boundary in the Minkowskian spacetime is $C^1$ at a point $p$ we will be able to compute the tangent vector to the singular boundary at $p$. Then, by definition, the causal character of the singular boundary at $p$ coincides with the causal character (spacelike, timelike or lightlike) of its tangent vector.
Furthermore, inspired by the definitions appearing in \cite{Penrose74}, we will also specify that there is a {\it past} spacelike (or {\it past} lightlike) singularity at $p$ if only past-directed causal curves end up at a spacelike (or lightlike) singularity at $p$ (see, for example, figures \ref{posib}(i) and \ref{posib}(iv), respectively). Likewise, we say that there is a {\it future} spacelike (or {\it future} lightlike) singularity at $p$ if only future-directed causal curves end up at a spacelike (or lightlike) singularity at $p$ (see, for example, figures \ref{posib}(ii) and \ref{posib}(v), respectively).

Let us now consider a sufficiently small simply connected open set $\mathcal U_M$ in the two-dimensional Minkowskian spacetime with corresponding points in the physical spacetime (specifically, in an open set $\mathcal U$ ---defined in section \ref{GSSM}---) and such that a part of its boundary consists of an open interval of the singular boundary containing $p$. Two nonvanishing null vector fields may be defined on $\mathcal U_M$ such that they are linearly independent at each point of $\mathcal U_M$ \cite{Beem}. The integral curves of the two null vector fields provide us with two families ($\mathcal{F}_1$ and $\mathcal{F}_2$) of (non-necessarily affine parametrized) null geodesics.

\subsection{Relating the causal character of singular points to null geodesics}

If we parametrize the $C^1$ curve describing the singular boundary with a parameter $\lambda$, then the value of the norm of its tangent vector will be given by a continuous function $\mathcal{N}(\lambda)$. The continuity of $\mathcal{N}(\lambda)$ implies that, if it is bigger than zero at a point $p$ corresponding to a certain $\lambda=\lambda_0$, then it will be bigger than zero for an open interval $\lambda_-<\lambda_0<\lambda_+$. So that there is an open interval around $p$ on the singular boundary where the boundary is necessarily spacelike. On the other hand, in this situation there are only two possibilities for the behaviour of the light-like geodesics at $p$ that traverse $\mathcal U_M$:
\begin{itemize}
\item There is a \textit{past (or future) spacelike singularity} at $p$ $\Rightarrow$ A null geodesic of every family \emph{leaves} (resp. reaches) $p$. (See fig. \ref{posib}-i (resp. fig. \ref{posib}-ii), where we have drawn the singularity horizontally\footnote{The situation does not change if we draw the singularity with an inclination bigger than -45º but less than 45º around $p$, as is guaranteed by $\mathcal{N}(\lambda_-<\lambda_0<\lambda_+)>0$.}.
\end{itemize}

Similarly,  if $\mathcal{N}(\lambda_0)<0$ then  $\mathcal{N}$ will be less than zero for an open interval $\lambda_-<\lambda_0<\lambda_+$ and there is only a possibility for the behaviour of the light-like geodesics at $p$ that traverse $\mathcal U_M$:
\begin{itemize}
\item There is a \textit{timelike singularity} at $p$ $\Rightarrow$ A null geodesic of one family \emph{reaches} $p$ and a null geodesic of the other family \emph{leaves} it. (See fig. \ref{posib}-iii).
\end{itemize}

It is useful to note that if one, and only one, null geodesic leaves or reaches a point $p$ in the boundary then, by process of elimination, the singularity must be lightlike at $p$. Specifically, there are two cases:
\begin{itemize}
\item[---] If only a null geodesic \emph{leaves} (or \emph{reaches}) $p$ then there is a \textit{past (resp. future) lightlike singularity} at $p$.  (An example\footnote{Note that, in general, the singularity does not have to be lightlike all around $p$.} of this situation is shown in fig. \ref{posib}-iv (resp. fig. \ref{posib}-v)).
\end{itemize}

\subsection{Intervals of piecewise constant causal characterization}

\begin{figure}
\includegraphics[scale=0.8]{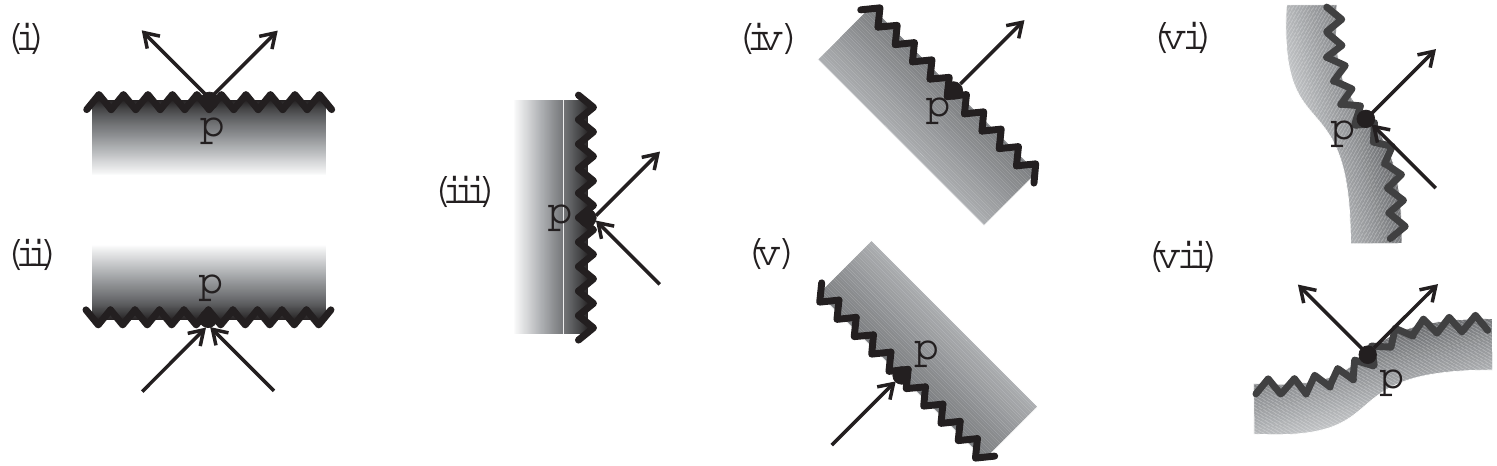}
\caption{\label{posib} The behaviour of null geodesics (arrows) at $p$ in an open $C^1$ interval of the singular boundary (wavy line) is shown for (i) a past spacelike singularity at $p$, (ii) a future spacelike singularity at $p$ and (iii) a timelike singularity at $p$. When only a null geodesic leaves or reaches $p$ there are two possibilities: (iv) $p$ is part of a past lightlike singularity or (v) $p$ is part of a future lightlike singularity. In (vi) there is a lightlike singularity at $p$ while the rest of the interval is timelike. Thus, it is an example of a \textit{piecewise timelike interval}. In (vii) there is a lightlike singularity at $p$ while the rest of the interval is spacelike. This is an example of a \textit{piecewise past spacelike interval}. }
\end{figure}

Let us now classify the intervals in the singular boundary according to the behaviour of the null geodesics. (We enumerate them according to the number of geodesics that leaves or reaches every of their points):
\begin{enumerate}
 \item[$1^-$ ($^+)\rangle$] Intervals where, for every of their points, only a null geodesic \emph{leaves} (resp. \emph{reaches}) it. Then, according to our previous subsection, the \emph{interval} is a \textit{past (resp. future) lightlike singularity}.
\item[$2^0\rangle$] Intervals where, for every of their points, a null geodesic of one family \emph{reaches} the point and a null geodesic of the other family \emph{leaves} it. Then we have seen in the previous subsection that the interval can contain points where it would be timelike ($\mathcal{N}(\lambda_0)<0$) and, as a result, subintervals around these points would be timelike.
    The possibility of having a \emph{point} in this singular interval where the singularity is spacelike is trivially discarded since the behaviour of the null geodesics does not correspond with the expected one according to the first two items in the previous subsection. The possibility of having a lightlike \emph{subinterval} is also discarded since, if we could draw the subinterval, it would have an inclination of $\pm 45º$, so that its points could not be reached by null geodesics from both families as demanded in this case. However, there is still the possibility for the interval to have lightlike \emph{points} only if \emph{every} neighborhood of the lightlike point contains timelike subintervals (an example is shown in figure \ref{posib}-vi).
     In this case we will say that the interval is \textit{piecewise timelike}\footnote{In fact, it can be argued that the distinction between strictly timelike and piecewise timelike intervals is not very relevant from a physical point of view. For example, if we are worried about the formation of a Naked Singularity in a particular model, we will have one in both cases.}.
\item[$2^-$ ($^+)\rangle$] Intervals where, for every of their points, a null geodesic of every family \emph{leaves} (resp. \emph{reaches}) it. Then we have seen in the previous subsection that the interval can contain points where it would be spacelike ($\mathcal{N}(\lambda_0)>0$)
    and, as a result, subintervals around these points would constitute a past (resp. future) spacelike singularity.
    Following the reasoning in the previous item, there is only the additional possibility of having lightlike \emph{points} if \emph{every} neighborhood of the lightlike point in the interval contains spacelike subintervals (an example is shown in figure \ref{posib}-vii (resp. its time reversal)). In this case we will say that the interval is a \textit{piecewise past (resp. future) spacelike singularity}.
\end{enumerate}

Here finishes our list of possible $C^1$ intervals according to the behaviour of the two families of null geodesics traversing $\mathcal U_M$. (Note that there cannot be $C^1$ \emph{intervals} such that every point is reached or left by a total of three or four null geodesics traversing $\mathcal U_M$).


\begin{figure}
\includegraphics[scale=1]{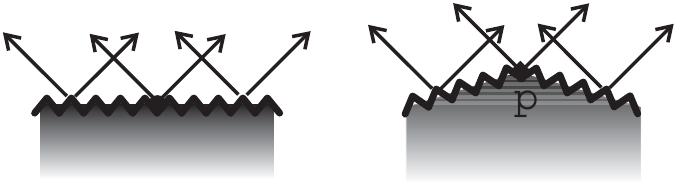}
\caption{\label{piecewiseint}
We show, to the left, a proper spacelike open interval of the singular boundary and, to the right, a piecewise spacelike singularity (since the -\textit{angular}- boundary it not $C^1$ at $p$). Both possibilities are indistinguishable if we only take into account that a null geodesic of every family leaves every point in the boundary.}
\end{figure}

As some exact solutions found in the literature show, we could also deal with just \textit{piecewise} $C^1$ boundaries. In these cases, there would be points where it will not be possible to define a tangent vector to the singular boundary and, strictly speaking, no matter the chosen approach to ascertain the causal character of the boundary we should resign ourselves to deal with \emph{piecewise causal characterizations}. An additional difficulty appears when one analyzes the null geodesics for these \textit{piecewise} $C^1$ boundaries in order to get its causal characterization. Consider, for example, an open interval in the singular boundary where two null geodesics leave the boundary at every of its points. Then we cannot tell with \emph{just} this scrutiny of null geodesics whether  there is some point on the interval where the singular boundary is not $C^1$ (see figure \ref{piecewiseint}).
%
%
Likewise, a point in the singular boundary where the boundary is not $C^1$ can pass unnoticed either in intervals where two null geodesics leave the boundary at every of its points (imagine the time reversal of figure \ref{piecewiseint}) or in intervals that are reached and left by one null geodesic of every family on every point. However, intervals where, at every point, only a null geodesic leaves it (past lightlike singularities) or reaches it (future lightlike singularities) are clearly $C^1$ (they are defined by $\pm45º$-inclined straight lines) and their causal character is therefore strictly determined.

%

\begin{figure}
\includegraphics[scale=0.8]{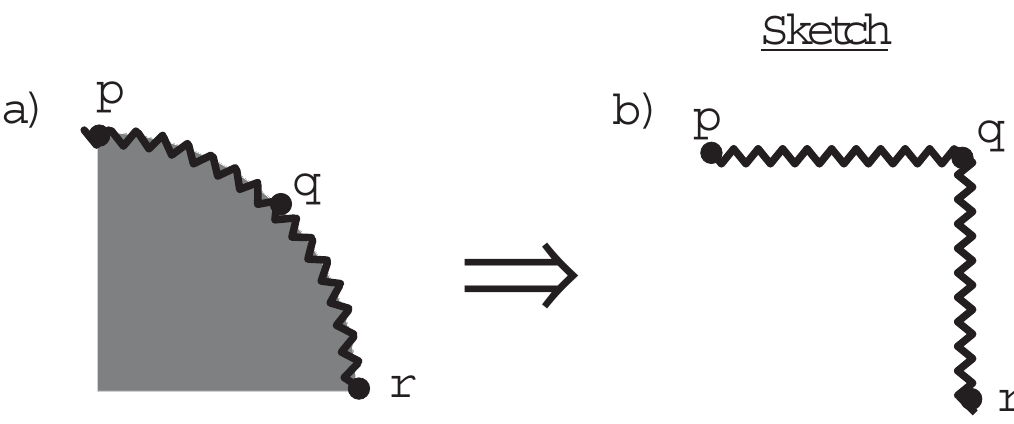}
\caption{\label{sketch}
In a) we draw an open interval of a singular boundary. In b) we show a sketch of its causal character. With a scrutiny of null geodesics we can infer that $(p,q)$ is a (piecewise) past spacelike interval and $(q,r)$ is a (piecewise) timelike interval. However, even if the boundary is lightlike at point $q$, we can not infer this with a simple scrutiny of null geodesics, since the same re-count (one reaches $q$ and two leave it) could be obtained if the boundary were $C^0$ at $q$. Therefore, in our sketch we overlook the possible lightlike character of the \textit{transition point} $q$.}
\end{figure}

Let us now call \textit{transition points} those points in the singular boundary where two different intervals from the ones defined above join (for example, a $2^0$ interval is followed by a $2^-$ interval in the singular boundary). This definition implies either that the singular boundary is lightlike at a transition point $p$ or that no tangent vector to the boundary can be defined at $p$. Since every piecewise $C^1$ boundary will be composed of intervals with a (piecewise) single causal character joined through transition points, we will be able to \emph{sketch} the causal character of the singular boundary by identifying the causal character of every (piecewise)-single-causal-character interval and overlooking whether the transition points really are lightlike. In order to exemplify this,
in figure \ref{sketch}) we have sketched a spacelike open interval joined to a timelike open interval through a \textit{transition point} $q$. In this spirit, from now on we will omit the adverb \textit{piecewise} in single-causal-character intervals.

\subsection{Translation into a physical spherically symmetric spacetime}

\begin{figure}
\includegraphics[scale=0.8]{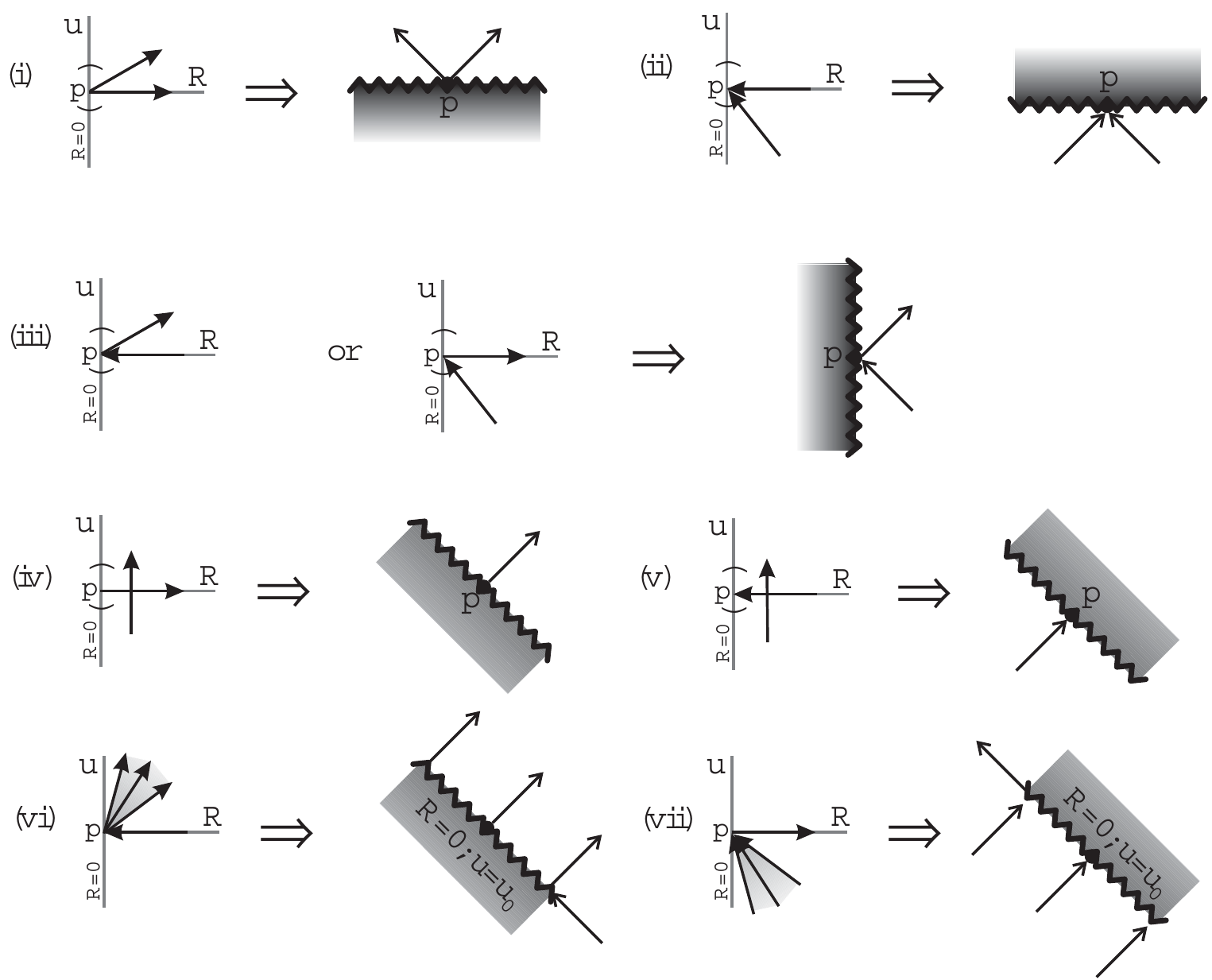}
\caption{\label{trans}
Translation table between the chart in the physical spacetime and the sketches in the unphysical spacetime for single causal character open intervals. In cases (i) to (iii) at every point in a whole open interval of $u$'s with $R=0$ a null geodesic of every family tends to (or departs from) the point. (We depict only those tending to (or departing from) a point $p$ $\{R=0,u=u_0\}$ in the interval). In cases (iv) and (v) only a family of null geodesics \emph{tend} to (or depart from) the points in a whole open interval of $u$'s with $R=0$. In the last possibilities, (vi) and (vii), an infinite number of null geodesics (in the grey zone of the $R-u$ diagram) \textit{tend} to a single point $\{R=0,u=u_0\}$, what must be interpreted as a lightlike singularity in the unphysical spacetime.}
\end{figure}

Let us assume now that we are working with a spherically symmetric spacetime and that we have identified the physical two-dimensional lorentzian surface orthogonal to the 2-spheres. Assuming that the invariant areal radius $R$ can be used as a coordinate in a local chart of the lorentzian surface, we will work in local coordinates $\{R,u\}$, where $u$ is a lightlike coordinate. Note that $u=$constant has a clear geometrical meaning, even if no further constraint is imposed on $u$, since it corresponds to an invariantly defined radial null geodesic. Moreover, according to our project, with this choice of coordinates a family of radial null geodesics is given ($u=$constant) and we only need to compute the other family of radial null geodesics in order to obtain the sketched causal characterization of any $R=0$ singularity. This is so because, as explained in the introduction, we know that for every null geodesic in this physical two-dimensional spacetime there is a corresponding null geodesic in the unphysical two-dimensional minkowskian spacetime and viceversa. The local translation from the physical ($\mathcal U$) to the unphysical ($\mathcal U_M$) spacetime is rather straightforward. We have listed the possibilities for single-causal-character open intervals in figure \ref{trans}. The only noteworthy case appears when an infinite number of null geodesics from the second family \textit{tends} to (or \textit{departs} from) a singularity at $\{R=0, u=u_0\}$, which corresponds to the existence of a lightlike singularity, as can be seen in figures \ref{trans}-(vi) and \ref{trans}-(vii).

%
%

\section{General spherically symmetric metric}\label{GSSM}
Let us consider an oriented four-dimensional spherically symmetric space-time
$\cal V$. In order to study the local causal behaviour around $R=0$ we will consider a local chart endowed with coordinates $\{x^{\mu}\}=\{u,R,\theta,\varphi\}$
($\mu=0,1,2,3$) and an open set $\mathcal{U}\equiv\{(u,R)| -\delta u< u< \delta u,\ 0< R< \delta R\}$.
The line-element can be expressed in this local chart
as
\begin{equation}
ds^2=-e^{4\beta}\chi du^2+2 \varepsilon e^{2\beta}\ du\ dR+R^2\
d\Omega^2 , \label{mI}
\end{equation}
where $R$ is the \textit{areal radius}, $\varepsilon^2=1$, $\beta$ and $\chi$
are assumed to be at least $C^{2-}$ functions\footnote{This is a minimum requirement which guarantees the existence and uniqueness of null geodesics \emph{in} $\mathcal{U}$ \cite{H&E}. It also enables to use Einstein's equations
(in case one works in the framework of GR).} on $\{u,R\}$ \emph{in} the local chart, $\beta$ is also assumed to be bounded as it approaches $R=0$, and \mbox{$d\Omega^2\equiv
d\theta^2+\sin^2\theta d\varphi^2$}.

In spherical symmetry one can define the \emph{scalar invariant}
\begin{displaymath}
m\equiv\frac{R}{2}(1-g^{\mu\nu}\partial_\mu R \partial_\nu R)
\end{displaymath}
(see \cite{M&S} and also \cite{Hayward}, \cite{PSR} and references therein).
If one computes $m$ for the metric (\ref{mI}) it is easily checked that
\begin{equation}
m=\frac{R}{2}(1-\chi). \label{graven}
\end{equation}

On the other hand, if we do not want $R=0$ scalar curvature singularities,
the scalar invariants polynomial in the Riemann tensor must remain finite at $R=0$.
It is well known that there are only four algebraically independent scalar invariants associated with a general spherically symmetric metric \cite{narkar}. We can take, for example, \cite{Carmi}\cite{Santo}\cite{Torres}:
\begin{eqnarray*}
\mathcal{R}&\equiv& g^{\alpha\gamma} g^{\beta\delta}
R_{\alpha\beta\gamma\delta},\\
r_1&\equiv&
{S_\alpha}^\beta {S_\beta}^\alpha,\\
r_2&\equiv&
{S_\alpha}^\beta {S_\beta}^\gamma
{S_\gamma}^\alpha,\\
w_2 &\equiv&
\bar{C}_{\alpha\beta\gamma\delta}
{\bar{C}^{\gamma\delta}}_{\ \ \mu\nu} \bar{C}^{\mu\nu\alpha\beta},
\end{eqnarray*}
where ${S_\alpha}^\beta \equiv {R_\alpha}^\beta-
{\delta_\alpha}^\beta \mathcal{R}/4$, being ${R_\alpha}^\beta$ the
Ricci tensor and $\mathcal{R}$ the curvature scalar;
$C_{\alpha\beta\gamma\delta}$ the Weyl tensor,
$\bar{C}_{\alpha\beta\gamma\delta}\equiv
(C_{\alpha\beta\gamma\delta} +i\ *C_{\alpha\beta\gamma\delta})/2$
is the complex conjugate of the selfdual Weyl tensor
being $*C_{\alpha\beta\gamma\delta}\equiv
\epsilon_{\alpha\beta\mu\nu} C^{\mu\nu}_{\ \ \gamma\delta}/2$ the dual of the Weyl tensor.
%
If one evaluates these invariants for (\ref{mI}) using (\ref{graven}) one arrives to
the following statement \cite{Torres}\cite{FST1}: All scalar invariants polynomial in the Riemann tensor will be finite at $R=0$, preventing the existence of scalar curvature singularities if,
and only if,
\begin{eqnarray}
\lim_{R\rightarrow 0} m = \lim_{R\rightarrow 0} \frac{m}{R}=\lim_{R\rightarrow 0} \frac{m}{R^2}=\lim_{R\rightarrow 0} \frac{\beta-\beta_0}{R}=0 ,\nonumber \\
\lim_{R\rightarrow 0}\frac{\beta-\beta_0}{R^2}=\beta
_2(u), \hspace{5mm} \lim_{R\rightarrow 0}\frac{m}{R^3}= m_3
(u),  \label{condisreg}
\end{eqnarray}
where $\beta _0 (u)\equiv \lim_{R\rightarrow 0}\beta (u,R)$ and both $\beta
_2(u)$ and $m_3 (u)$ are finite functions.

Moreover, if the set of scalar invariants is finite at $R=0$ there will be only two algebraically independent scalar invariants at $R=0$, say $\mathcal{R}$ and $r_1$, since
$w_2=0$ and $343 {r_1}^6=3087 {r_2}^4$.

Let us finally remark that one of the interesting properties of the scalar curvature singularities is that their existence can not be an artifact of the coordinate system used.
We now know that if conditions (\ref{condisreg}) are not fulfilled then for any curve approaching $R=0$ there will be at least one scalar invariant that will grow without limit along the curve as it approaches the $R=0$-singularity. It is obvious that, if one uses different local coordinates, the value of the \emph{scalar invariant} at every point of the curve must coincide, when computed with the new local coordinates, with the values obtained at the same points with the coordinate system used in this article. Therefore, no matter what local coordinates are used, the same scalar invariant will grow without limit along the curve, what unmistakably identifies a $R=0$-scalar curvature singularity.


%


\subsection{Radial null geodesics}\label{nullgeodesics}
We choose $u$ growing to the future. Then
\begin{equation}
\mathbf{l}=\frac{d}{d \ell}=- \frac{\varepsilon}{2 R} \frac{\partial}{\partial R} \label{deflv}
\end{equation}
is a future directed radial null vector tangent to the null geodesics $u=$constant (namely, the $\mathcal{F}_1$ family) and $\ell$ is a future directed parameter.
Since the expansion \cite{H&E} of these null geodesics is given by
\begin{displaymath}
\theta_l=\frac{-\varepsilon}{R^2},
\end{displaymath}
if $\varepsilon=-1$ (or $+1$), the expansion
is positive (negative, respectively) and, they are \emph{outgoing} (\emph{ingoing}, respectively) radial null geodesics directed towards increasing $R$'s (decreasing $R$'s, respectively) according to (\ref{deflv}).
Therefore, the behaviour of the $\mathcal{F}_1$ family, with regard to whether the null geodesics are coming or are directed towards the $R=0$-singularity, is absolutely defined by the sign of $\varepsilon$, a fact that we will use throughout this article.
On the other hand,
\begin{equation}
\mathbf{k}=\frac{d}{d \kappa}= 2 R e^{-2 \beta} \frac{\partial}{\partial u}+ \varepsilon (R-2 m) \frac{\partial}{\partial R} \label{defkv}
\end{equation}
is a future directed radial null vector such that $\mathbf{l}\cdot\mathbf{k}=-1$ and $\kappa$ is a future directed parameter. For later purposes let us write explicitly the equations governing the geodesics that have $\mathbf{k}$ as its tangent vector field (the $\mathcal{F}_2$ family):
\begin{equation}
\left\{
\begin{array}{l}
 \frac{dR}{d\kappa}= \varepsilon [R-2 m(u,R)] \label{system}\\
\frac{du}{d\kappa}= 2 R e^{-2\beta}
\end{array}
\right.
\end{equation}
In this case the expansion is given by
\begin{displaymath}
\theta_k=\frac{2 \varepsilon}{R} (R-2 m).
\end{displaymath}
If at a given 2-sphere $\chi>0$ ($\Leftrightarrow R>2 m$) and $\varepsilon=-1$ (or $+1$), $\mathbf{k}$ is tangent to a family of null geodesics with negative (positive, respectively) expansion, the areal coordinate $R$ decreases (increases, respectively) along them according to (\ref{system}) and, therefore, these radial null geodesics are \emph{ingoing} (\emph{outgoing}, respectively) in the considered 2-sphere.
However, it is interesting to note
that if, in a given 2-sphere, $\chi<0$ and $\varepsilon=-1$
($\varepsilon=+1$) then the two radial null vectors have both positive
(negative) expansion which means that the 2-sphere is a closed surface trapped to its past (future, respectively).

\section{Characterization of the singularity: Case $\lim_{\kappa\rightarrow \kappa_0} m(u,R)\neq 0$}\label{csmneq0}

Let us consider, in $\mathcal{U}$, the set of radial null geodesics belonging to the family $\mathcal{F}_2$ that approach $R=0$ (either toward their past or their future). For every such geodesic we define $\kappa_0$ to be the value of its parameter such that  $\lim_{\kappa\rightarrow \kappa_0} (R(\kappa),u(\kappa)) = (0,u_0)$, where the precise value $u_0 \in (-\delta u,\delta u)$ depends on every geodesic.
Now, if $\lim_{\kappa\rightarrow \kappa_0} m(u,R)\neq 0$ along every of these null geodesics, according to (\ref{system}),
\begin{eqnarray}
\lim_{\kappa\rightarrow \kappa_0} \frac{dR}{d\kappa}&=&-2 \varepsilon \lim_{\kappa\rightarrow \kappa_0} m(u,R)\label{mneq0}\\
\lim_{\kappa\rightarrow \kappa_0} \frac{du}{d\kappa}&=&0\nonumber
\end{eqnarray}

What implies that:
\begin{itemize}
\item If $\lim_{\kappa\rightarrow \kappa_0} m(u,R)> 0$ in $\mathcal U$ then the two different families of radial null geodesics are outgoing ($\varepsilon=-1$) or ingoing ($\varepsilon=+1$ ) in $\mathcal U$ (see sect.\ref{nullgeodesics} for the $u=$constant family) and, therefore, the singularity is spacelike at $R=0$.
\item If $\lim_{\kappa\rightarrow \kappa_0} m(u,R)< 0$ in $\mathcal U$ then one radial null geodesic is outgoing while the other is ingoing and, therefore, the singularity is timelike at $R=0$.
\end{itemize}

As a corollary we have the following more workable and well-known \cite{Hayward} result:  if $\lim_{(u\rightarrow u_0,R\rightarrow 0)} m(u,R)$ exits and it is positive for all $u_0 \in \mathcal U$ then the singularity is
spacelike at $R=0$ and
if $\lim_{(u\rightarrow 0,R\rightarrow 0)} m(u,R)$ exists and it is negative for all $u_0 \in \mathcal U$ then the singularity is
timelike at $R=0$.

\section{Characterization of the singularity: Case $\bar{m}(u=0,R=0)=0$ and $\bar{m},_u(u=0,R=0)\neq0$}\label{m0hyp}

In this section we will assume, as usual, that $m$ and $\beta$ are $C^{2-}$ in $\mathcal U$ and also that $C^1$ auxiliary extensions (devoid of any physical meaning) $\bar m$ and $\bar \beta$, respectively, exist in $\bar \mathcal{U} \equiv\{(u,R)| -\delta u< u< \delta u,\ -\delta R < R< \delta R\}$
such that, formally,
$\bar m= m$ and $\bar \beta=\beta$ for  $(-\delta u< u< \delta u, 0 < R< \delta R)$\footnote[1]{Note that in the literature one usually finds the same functions $m$ and $\beta$ used for $R\leq 0$, what could be considered as the \textit{natural} extension.}.

In the case considered in this section ($\bar{m}(u=0,R=0)=0$ and $\bar{m},_u(u=0,R=0)\neq0$) there is a scalar curvature singularity at $R=u=0$ since the regularity conditions (\ref{condisreg}) are violated.

Without loss of generality, we will consider from now on
$\bar{\beta}(u,0)=0$ since, if it was not, we can always perform a coordinate change $u \rightarrow u'$ such that the new coordinate $u'$ were defined by $du'= e^{2 \bar{\beta}(u,0)} du$.

Now $u=R=0$ is a critical point of system (\ref{system}). In order to analyze the qualitative behaviour of the radial null geodesics we try a linear approximation at the critical point:
If we define $f^R\equiv\varepsilon [R-2 \bar{m}(u,R)]$ and $f^u\equiv 2 R e^{-2\bar{\beta}}$, then the \textit{linearization matrix}
is $A^\alpha_\beta\equiv f^\alpha,_\beta(u=0,R=0)$:
\begin{equation}\label{Amatrix}
A=\left(
\begin{array}{cc}
 \varepsilon [1-2 \bar{m},_R(0,0)] & -2\varepsilon \bar{m},_u(0,0) \\
 2 & 0
\end{array}
\right)
\end{equation}
The characteristic roots for this matrix are:
\begin{equation}
\lambda_\pm=\frac{\varepsilon(1-2 \bar{m},_R(0,0))\pm\sqrt{(1-2 \bar{m},_R(0,0))^2-16 \varepsilon \bar{m},_u(0,0)}}{2}
\end{equation}
The critical point $u=R=0$ will be \textit{hyperbolic} \cite{Perko} if none of the characteristics roots have zero real part.
In order to analyze this for our case, let us define $\Delta\equiv(1-2 \bar{m},_R(0,0))^2-16 \varepsilon \bar{m},_u(0,0)$. Then $u=R=0$ will be hyperbolic if
\begin{equation}
\bar{m},_u(0,0)\neq0
\end{equation}
and, in case $\Delta\leq0$, if the extra-condition $\bar{m},_R(0,0)\neq1/2$ is satisfied.
Under the assumption that $\bar m$ and $\bar \beta$ are $C^1$ we can apply the \textit{Hartman-Grobman Theorem} \cite{Perko}. According to this theorem, if the critical point is hyperbolic
then the behaviour of the nonlinear system near the critical point is qualitatively determined by the behaviour of the linear system $\dot{x}^\alpha=A^\alpha_\beta x^\beta$. Furthermore, the theorem implies that the qualitative behaviour is independent of the extension chosen for $m$ and $\beta$
when $\{-\delta u< u< \delta u, -\delta R < R\leq 0\}$ as long as it is a $C^1$ extension. In other words, and taking into account (\ref{Amatrix}), the results will only depend on the partial derivatives of $m$ as we approach $R=u=0$.
Note that in the hyperbolic case and near the critical point the curves defined by $f^R(u,R)=0$ and $f^u(u,R)=0$ behave approximately as straight lines in the $\{ u, R \}$ plane intersecting at the point $u=R=0$, which is an \textit{isolated} critical point \cite{Andronov}.
\subsection{Different possibilities}
\begin{itemize}
\item If $\Delta<0$ and $\bar{m},_R(0,0)\neq 1/2$ the characteristic roots are complex conjugate of each other with nonzero real part. This means that the critical point $u=R=0$ is a \textit{focus} \cite{Perko}\cite{Nemy}. In this way, the \emph{future directed}\footnote{Throughout the article we will use the fact that the radial null geodesics are future directed in order to interpret in correct physical terms the results provided by the qualitative theory of dynamic systems.} radial null geodesics in a neighborhood of the critical point must start at $R=0$, $u<0$ and end at $R=0$, $u>0$ after going around the critical point. Therefore,
    \begin{itemize}
    \item If $\varepsilon=-1$ then
    the $R=0$ singularity must be spacelike near the critical point for $u<0$
    while it must be timelike near the critical point for $u>0$. See figure \ref{center-focus}.
    \item If $\varepsilon=+1$ then
    the singularity must be timelike near the critical point for $u<0$
    while it must be spacelike near the critical point for $u>0$.
    \end{itemize}
\item If $\Delta<0$ and $\bar{m},_R(0,0)=1/2$ the point is not hyperbolic. However, even if the linearization does not suffice to distinguish the exact qualitative behaviour it can be guaranteed that $u=R=0$ will be either a focus or a center \cite{Perko} \cite{Jordan}. Therefore, the same characterization than in the above item applies.

\begin{figure}
\includegraphics[scale=0.8]{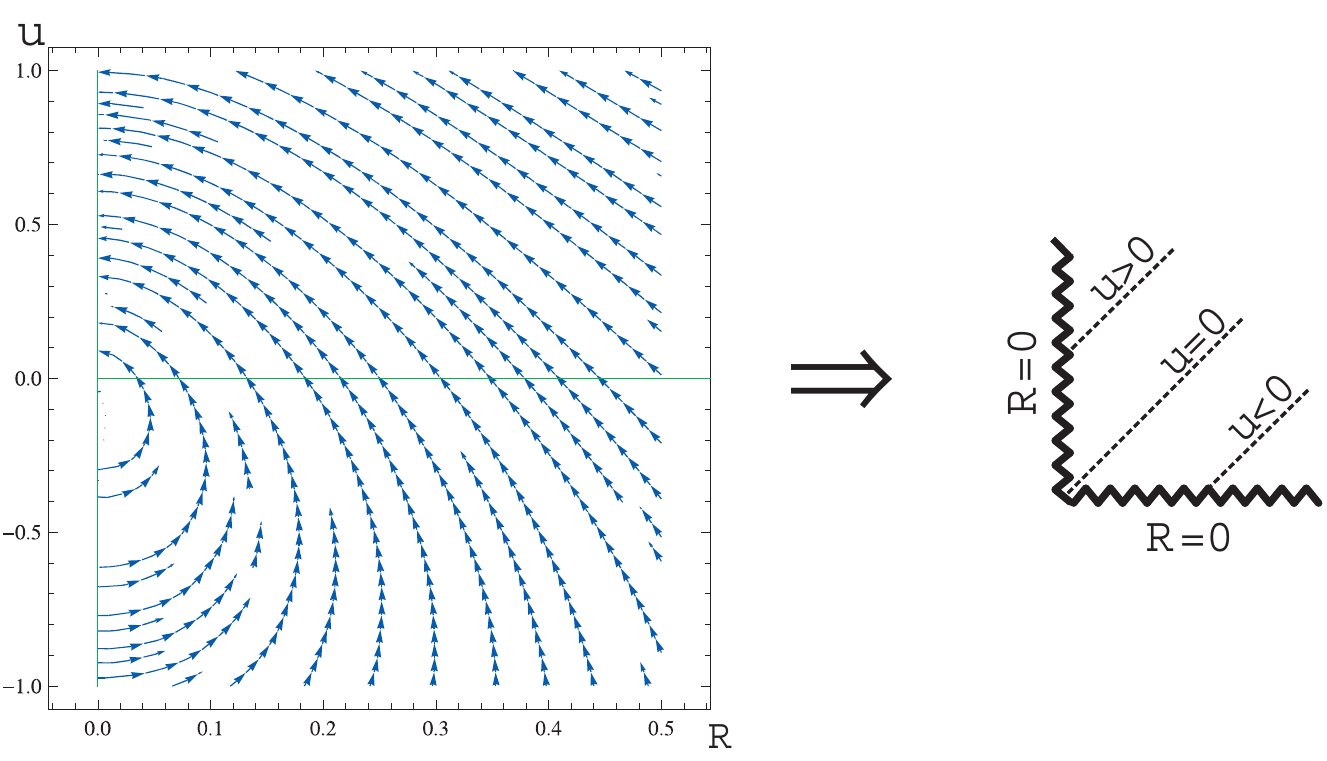}
\caption{\label{center-focus} In the left of this figure we sketch the \textit{center-focus} behaviour of the
second family of radial null geodesics
in the $\varepsilon=-1$ case near $u=R=0$ when det$A>0$ and $\Delta<0$. (In fact, it is only a half of the \textit{center-focus} behaviour since we are only interested in $R\geq0$). Note that the sketchy representation gives rise to the illusory discontinuousness of some curves. In the right we show the corresponding sketched Penrose diagram around $u=R=0$ with its characteristic spacelike-timelike behaviour in the $\varepsilon=-1$ case.}
\end{figure}

\item If $\Delta>0$ and $\varepsilon \bar{m},_u(0,0)>0$ then the characteristic roots are real, with the same sign and distinct so that the critical point is a \textit{node} \cite{Perko}\cite{Nemy}. The slope of the radial null geodesics ending or starting at $u=R=0$ is  $\xi_0 \equiv \lim_{\kappa\rightarrow \kappa_0} u(\kappa)/R(\kappa)$. Using (\ref{system}) we find $\xi_0= 2\varepsilon /(1-2\{\bar{m},_R (0,0) +\xi_0 \bar{m},_u(0,0)\})$. The two real roots $\xi_{0\pm}$ of this quadratic equation are

    \begin{equation}\label{crdir}
    \xi_{0\pm}=\frac{(1-2 \bar{m},_R(0,0))\pm\sqrt{(1-2 \bar{m},_R(0,0))^2-16 \varepsilon \bar{m},_u(0,0)}}{4 \bar{m},_u(0,0)}.
    \end{equation}

    \begin{itemize}
    \item If $T\equiv\varepsilon(1-2 \bar{m},_R(0,0))>0$ then $\lambda_\pm>0$, $\xi_{0\pm}>0$ and the node is unstable. Then it can be shown \cite{Jordan} that around $u=R=0$ all, except for one, radial null geodesics must start at this point with a definite slope which is $\xi_{0+}$ for $\varepsilon=+1$ and $\xi_{0-}$ for $\varepsilon=-1$. The exception is just one null geodesic starting with slope $\xi_{0-}$ for $\varepsilon=+1$ and $\xi_{0+}$ for $\varepsilon=-1$. The behaviour of the second family of null geodesics ($\mathcal{F}_2$) for the $\varepsilon=-1$ case together with the corresponding sketched Penrose's diagram is shown in figure \ref{unstable}.
\begin{figure}
\includegraphics[scale=0.8]{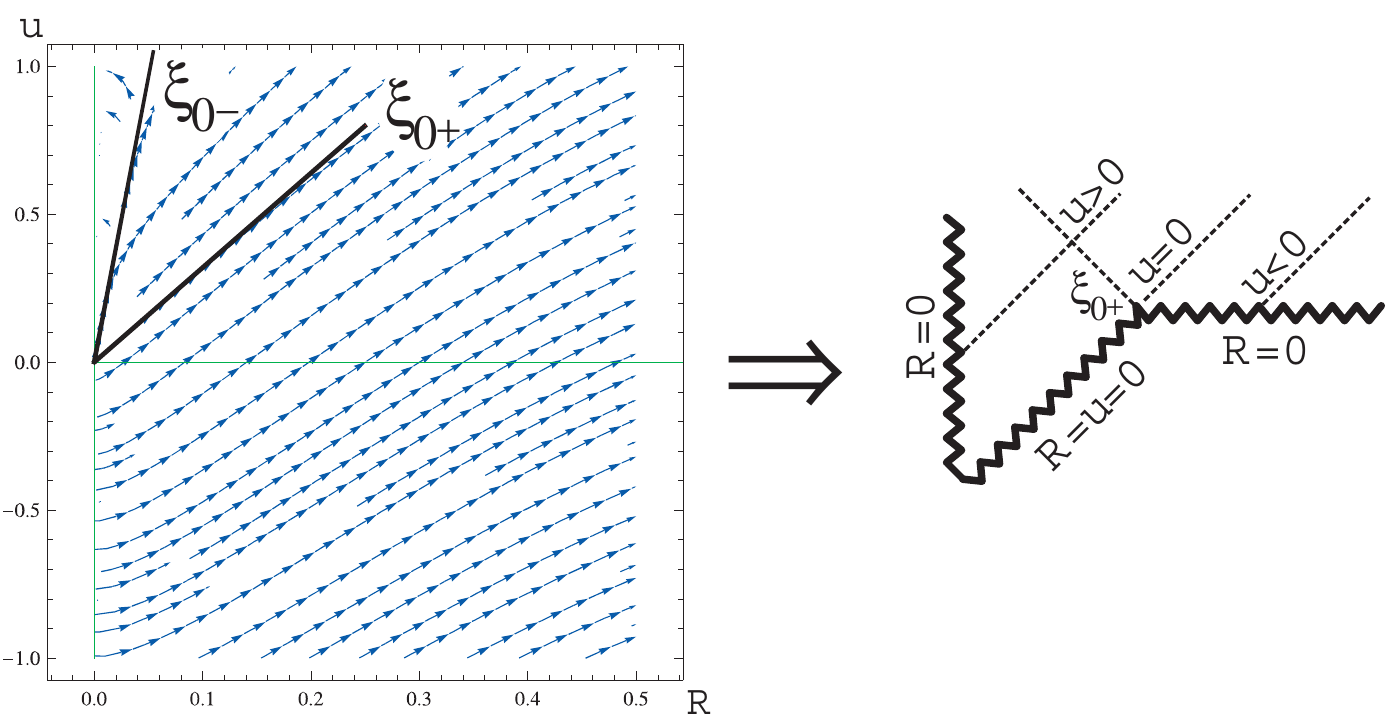}
\caption{\label{unstable} In the left of this figure we sketch the \textit{unstable node} behaviour of the second family of radial null geodesics
in the $\varepsilon=-1$ case near $u=R=0$ when det$A>0$, $T>0$ and $\Delta\geq 0$. The two straight lines crossing at $u=R=0$ indicate the critical directions of the radial null geodesics starting at $u=R=0$. From these geodesics only one leaves the critical point with direction $\xi_{0+}$ while all the others leave it with direction $\xi_{0-}$. In the right we show the corresponding Penrose diagram around $R=u=0$ with its characteristic $R=u=0$-lightlike singularity. Here we have pointed out the single radial null geodesic that leaves the critical point in the direction $\xi_{0+}$.}
\end{figure}

    \item If $T<0$ then $\lambda_\pm<0$, $\xi_{0\pm}<0$ and the node is stable. The comments for this case are similar to the case above. We summarize the behaviour of the second family of null geodesics for the $\varepsilon=-1$ case together with the corresponding sketched Penrose's diagram in figure \ref{stable}.
\begin{figure}
\includegraphics[scale=0.8]{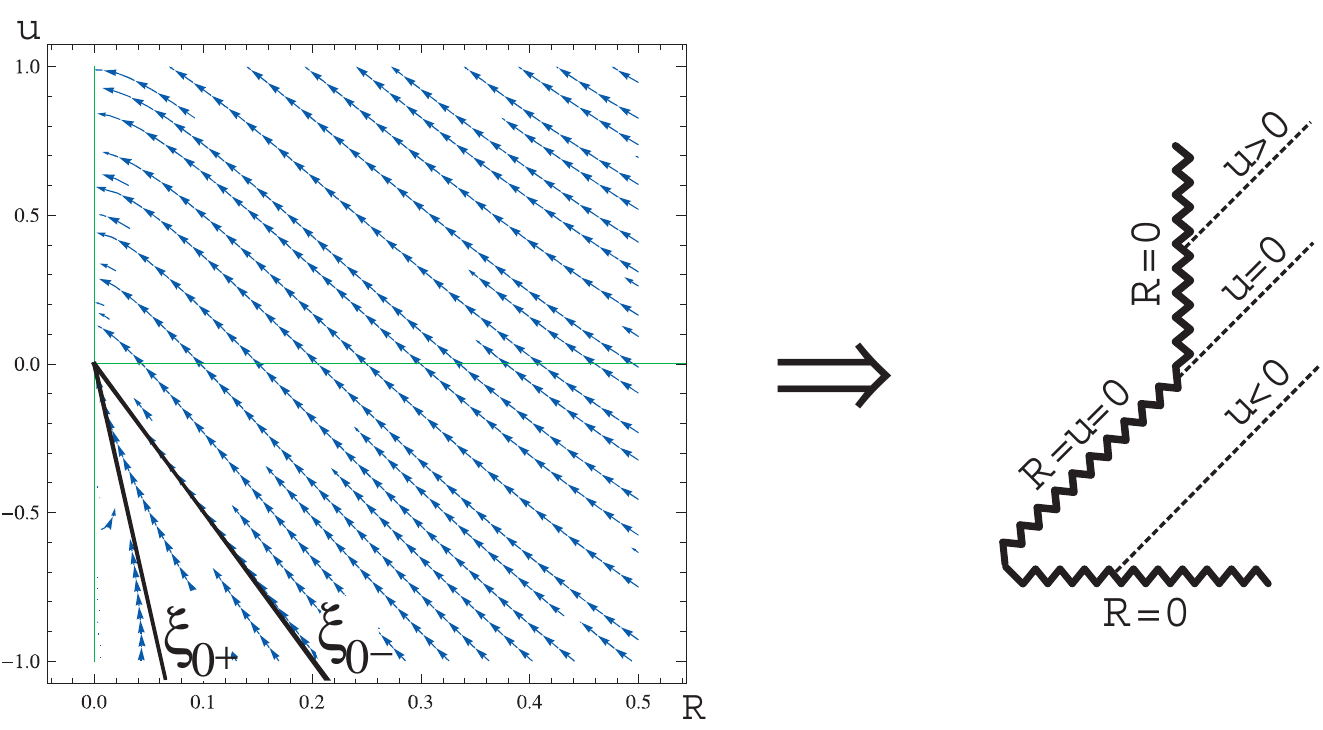}
\caption{\label{stable} In the left of this figure we sketch the \textit{stable node} behaviour of the second family of radial null geodesics
in the $\varepsilon=-1$ case near $u=R=0$ when det$A>0$, $T<0$ and $\Delta\geq 0$. In the right we show the corresponding Penrose diagram around $R=u=0$ with its characteristic $R=u=0$-lightlike singularity.}
\end{figure}
    \end{itemize}

\item If $\Delta=0$ and $\varepsilon \bar{m},_u(0,0)>0$ then the characteristic roots are real and equal ($\lambda$) so that the critical point is a \textit{degenerate node} \cite{Perko}\cite{Nemy}. There also exists an only critical direction ($\xi_{0}$).

    \begin{itemize}
    \item If $T>0$ then $\lambda>0$, $\xi_{0}>0$ and the node is unstable. Then it can be shown \cite{Jordan} that around $u=R=0$ all radial null geodesics must start at this point with a definite slope which is $\xi_{0}$ for $\varepsilon=\pm 1$.
        Therefore, the result is similar to figure \ref{unstable} with equal critical directions.

    \item If $T<0$ then $\lambda<0$, $\xi_{0}<0$ and the node is stable.
        Therefore, the result is similar to figure \ref{stable} with equal critical directions.
    \end{itemize}

\item If $\Delta>0$ and $\varepsilon \bar{m},_u(0,0)<0$ (no matter the value of $\varepsilon(1-2 \bar{m},_R(0,0))$) the characteristics roots are real with opposite sign and the critical point is a \textit{saddle} \cite{Perko}\cite{Nemy}. Likewise, the two critical directions also have opposite sign since $\mbox{sign}(\xi_{0-})=\varepsilon=-\mbox{sign}(\xi_{0+})$. The behaviour of the second family of radial null geodesics for the $\varepsilon=-1$ case together with the corresponding sketched Penrose's diagram is shown in figure \ref{saddle}.
\begin{figure}
\includegraphics[scale=0.7]{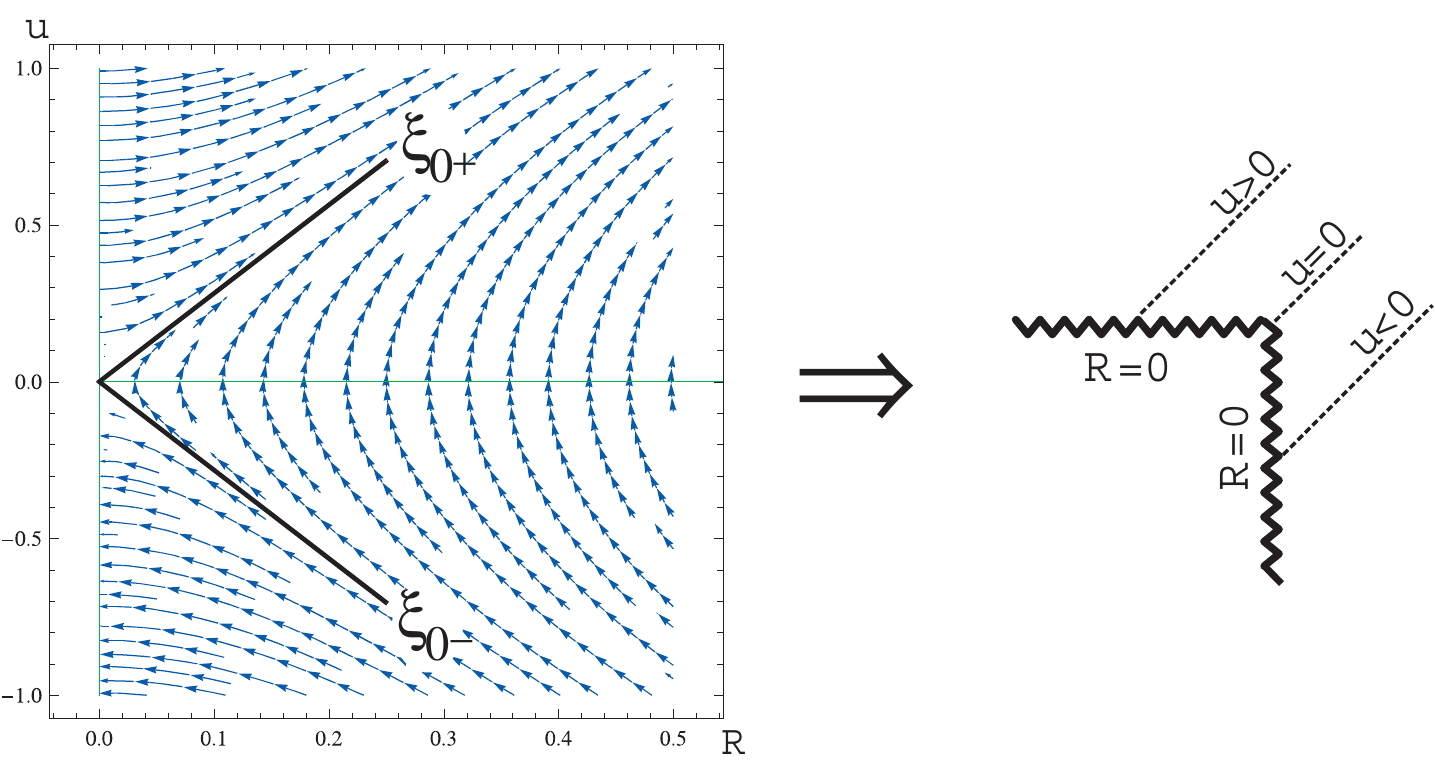}
\caption{\label{saddle} In the left of this figure we sketch the \textit{saddle} behaviour of the second family of radial null geodesics
in the $\varepsilon=-1$ case near $u=R=0$ when det$A<0$ and $\Delta>0$. In the right we show the corresponding sketched Penrose diagram around $u=R=0$ with its characteristic timelike-spacelike for the $\varepsilon=-1$ case.}
\end{figure}
\end{itemize}

\begin{figure}
\includegraphics[scale=0.7]{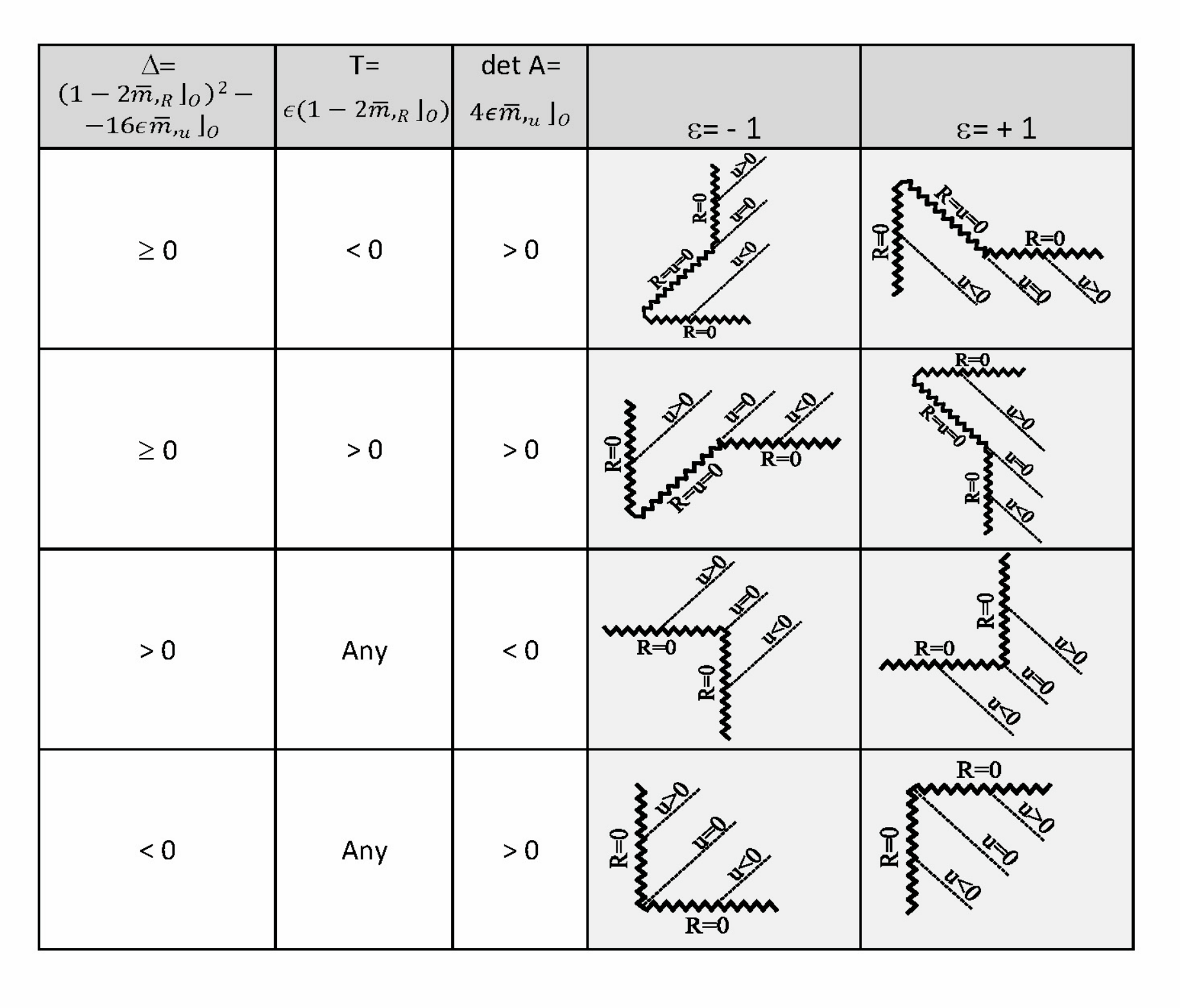}
\caption{\label{hyper} Characterization of the singularity when $\bar{m}(u=0,R=0)=0$ and $\bar{m},_u(u=0,R=0)\neq0$.}
\end{figure}

We have collected the results for $\bar{m},_u(u=0,R=0)\neq0$ in figure \ref{hyper}.

\section{Characterization of the singularity: Case $\bar{m}(u=0,R=0)=0$ and $\bar{m},_u(u=0,R=0)=0$}\label{m0nonhyp}

In this case
$u=R=0$ is an isolated critical point of system (\ref{system}), but, as we showed in the previous section, since $\bar{m},_u(u=0,R=0)=0$ the critical point is not hyperbolic. In fact, if we demand $\bar{m},_R(0,0)\neq 1/2$ the critical point will be \textit{semi-hyperbolic} \cite{DLLA} and we will have to use a different approach. (The theory regarding the qualitative behaviour of these points can be found in \cite{Andronov}\cite{DLLA}. The reader can also find a summary of the main results in appendix A). In order to apply the theory for semi-hyperbolic points we must now demand, first, the existence of a natural number $n\geq 2$ such that $\lim_{(u\rightarrow 0,R\rightarrow 0)} \partial^n m/\partial u^n (u,R)\neq 0$, while $\lim_{(u\rightarrow 0,R\rightarrow 0)} \partial^i m/\partial u^i (u,R)= 0$ for $i=1,...,n-1$, and, second, of $C^n$ extensions\footnote{Private communication with the authors of \cite{DLLA}, where $C^\infty$ is assumed independently of the value of the finite $n$.} $\bar{m}$ and $\bar{\beta}$.

First we define
\begin{eqnarray}
x&=& - \frac{2 \varepsilon}{1-2 \bar{m},_R(0,0)} R + u\\
y&=& R\\
t&=& \varepsilon (1-2 \bar{m},_R(0,0)) \kappa.
\end{eqnarray}
Then the system (\ref{system}) can be rewritten in the normal form
\begin{equation*}
\left\{
\begin{array}{l}
 \frac{dx}{dt}= P_2(x,y)\\
\frac{dy}{dt}= y+ Q_2(x,y),
\end{array}
\right.
\end{equation*}
where $P_2(x,y)\equiv \tilde{P}_2(u(x,y),R(y))$, $Q_2(x,y)\equiv \tilde{Q}_2(u(x,y),R(y))$,
\begin{eqnarray*}
         \tilde{P}_2(u,R)=\frac{2 \varepsilon}{1-2 \bar{m},_R(0,0)}\left( R e^{-2 \beta(u,R)} - \frac{R-2 \bar{m}(u,R)}{1-2 \bar{m},_R(0,0)}\right),\\
         \tilde{Q}_2(u,R)= 2\frac{\bar{m},_R(0,0) R- \bar{m}(u,R) }{1-2 \bar{m},_R(0,0)}
\end{eqnarray*}
and they satisfy
$P_2(0,0)=Q_2(0,0)=P_2,_x(0,0)=Q_2,_x(0,0)=P_2,_y(0,0)=Q_2,_y(0,0)=0$.
The assumed degree of differentiability together with the implicit function theorem guarantees that the equation $y+ Q_2(x,y)=0$ has a solution $y=\varphi(x)$ which can be written (if one considers the Taylor polynomial for $\bar{m}(u,R)$ and does some algebra) as the truncated series expansion\footnote{Note that, the requirement on
the degree of differentiability for $\bar m$ and $\bar{\beta}$ --taking into account Taylor's theorem-- allows us to write some functions derived from $\bar m$ and $\bar{\beta}$ as a Taylor polynomial plus a remainder term. From now on we will only write the first non-zero term of the Taylor polynomial (the only that is guaranteed to exist thanks to our assumptions) and suspension points --as in (\ref{yh})--.}
\begin{equation}
y=\frac{2}{n!} \frac{\partial^n \bar{m}/\partial u^n(0,0)}{1-2 \bar{m},_R(0,0)} x^n + ...,\label{yh}
\end{equation}
where $n\geq 2$ is assumed to be finite. If we define the function $\psi(x)=P_2(x,\varphi(x))$ then its truncated series expansion will have the form
\begin{equation}
\psi(x)= \Delta_n x^n+...,
\end{equation}
 where, in our case,
\begin{equation}
\Delta_n= \varepsilon c^2  \frac{\partial^n \bar{m}}{\partial u^n}(0,0),\label{Delta}
\end{equation}
and $c^2\equiv 4/(n! (1-2 \bar{m},_R(0,0))^2)>0$.

\begin{figure}
\includegraphics[scale=0.7]{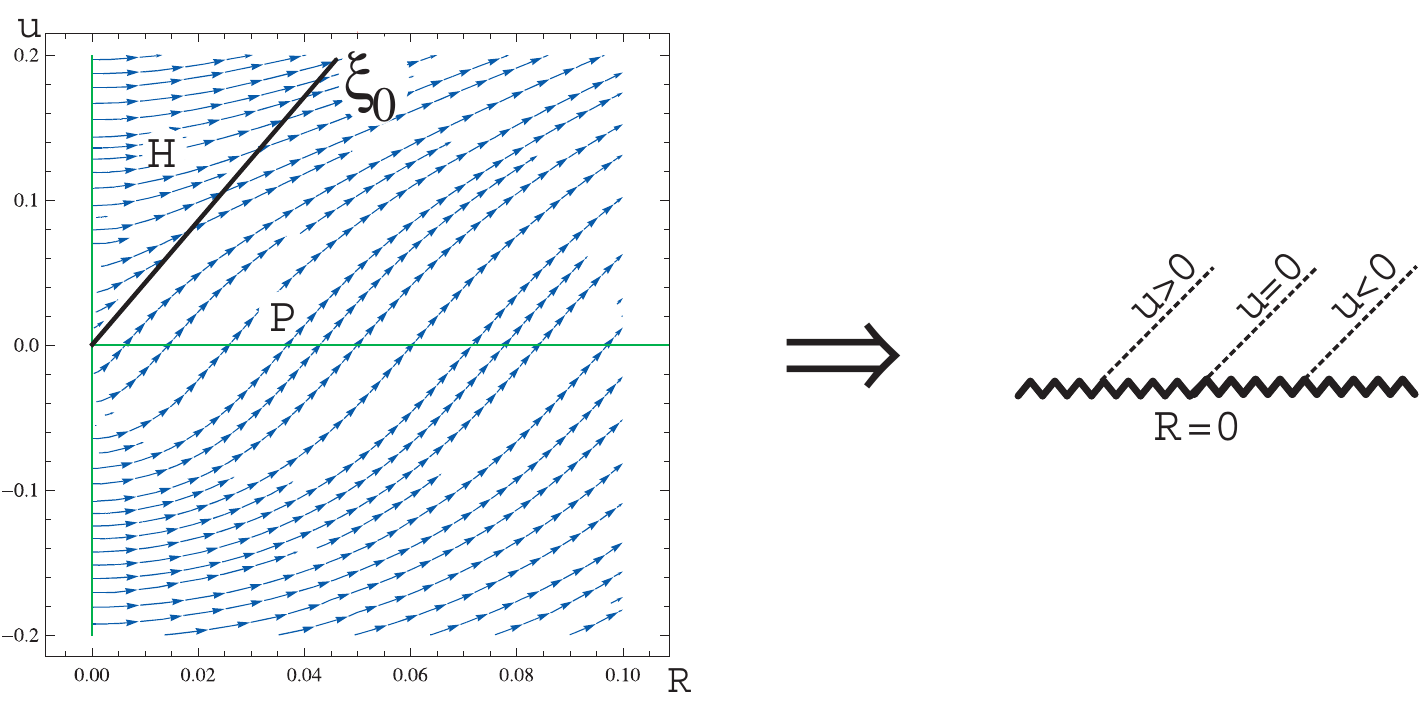}
\caption{\label{space} In the left of this figure we sketch the behaviour of the second family of radial null geodesics
in the $\varepsilon=-1$ case near $u=R=0$ when $\bar{m},_u(0,0)=0$, $\Delta_n<0$ with $n$ even and $\bar{m},_R(0,0)>1/2$. Only the $R\geq0$ part of the \textit{saddle node} structure is shown. The straight line marks the regular critical direction $\xi_0$ \cite{Nemy} for the curve starting at $u=R=0$ that separates a hyperbolic sector (H) from the parabolic sector (P). In the right we show the corresponding Penrose diagram around $u=R=0$ with its spacelike singularity.}
\end{figure}

\begin{figure}
\includegraphics[scale=0.7]{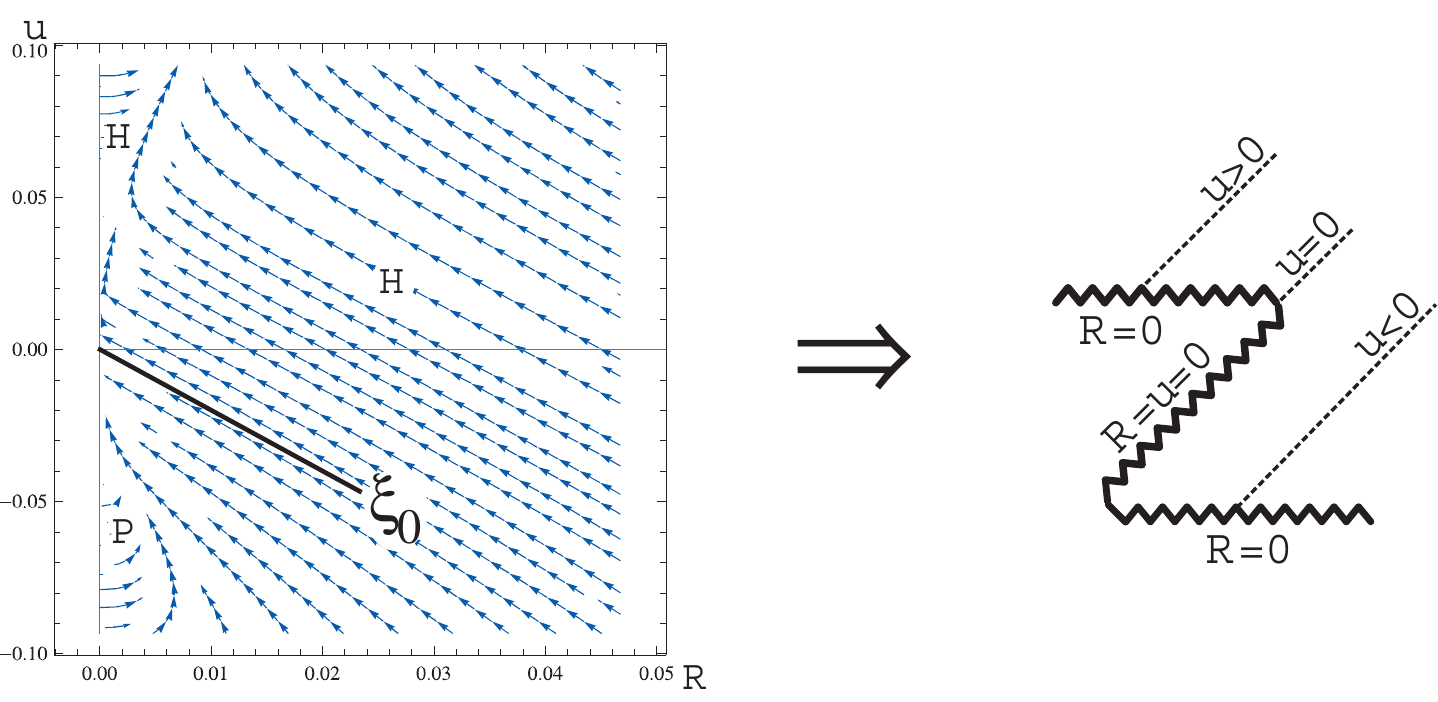}
\caption{\label{spacelight} In the left of this figure we sketch the behaviour of the second family of radial null geodesics
in the $\varepsilon=-1$ case near $u=R=0$ when $\bar{m},_u(0,0)=0$, $\Delta_n<0$ with $n$ even and $\bar{m},_R(0,0)<1/2$. We also show by means of a straight line the regular critical direction $\xi_0$. In this case the two hyperbolic sectors (H) and the parabolic sector (P) are partially visible for $R\geq0$. In the right we show the corresponding Penrose diagram around $u=R=0$ with its lightlike singularity.}
\end{figure}

\begin{figure}
\includegraphics[scale=0.7]{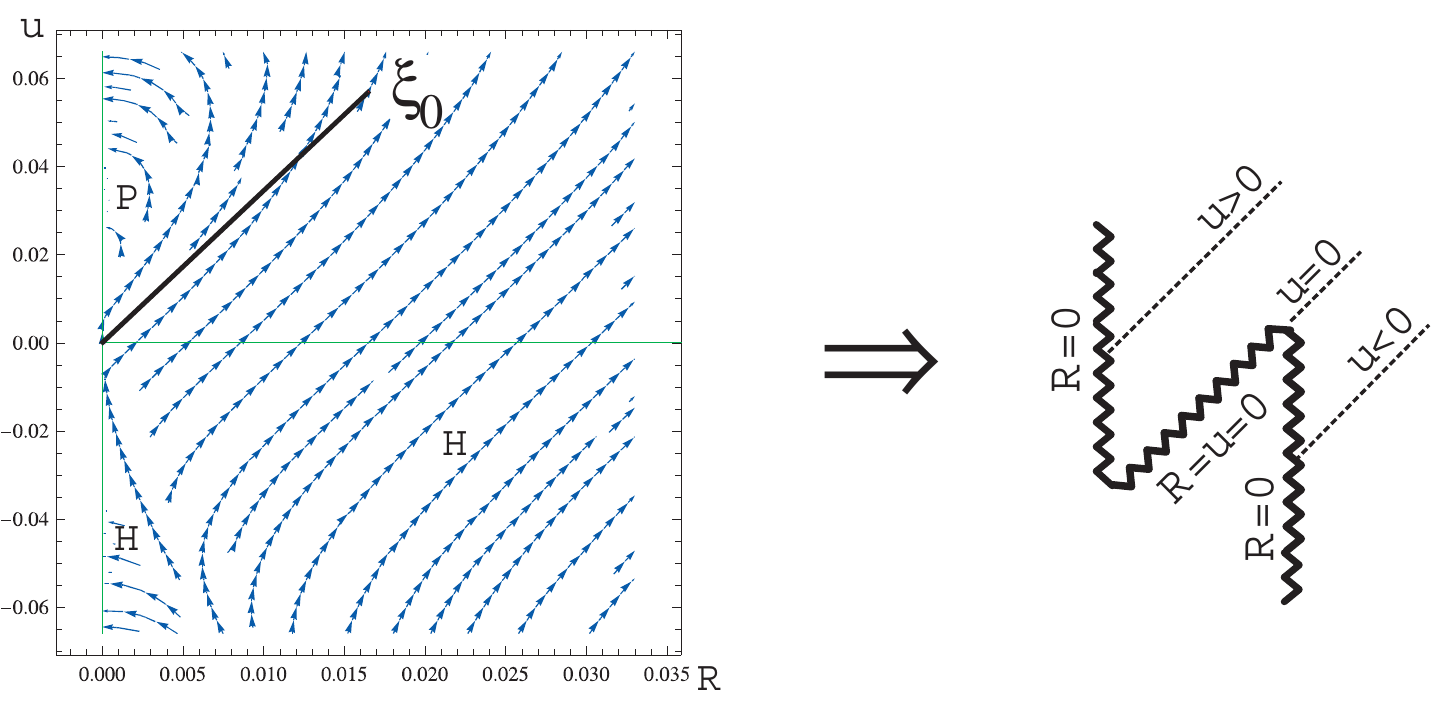}
\caption{\label{timelight} In the left of this figure we sketch the behaviour of the second family of radial null geodesics
in the $\varepsilon=-1$ case near $u=R=0$ when $\bar{m},_u(0,0)=0$, $\Delta_n>0$ with $n$ even and $\bar{m},_R(0,0)>1/2$. The two hyperbolic sectors (H) and the parabolic sector (P) are partially visible for $R\geq0$. We also show by means of a straight line the finite critical direction $\xi_0$. In the right we show the corresponding Penrose diagram around $u=R=0$, which is a lightlike singularity.}
\end{figure}

\begin{figure}
\includegraphics[scale=0.7]{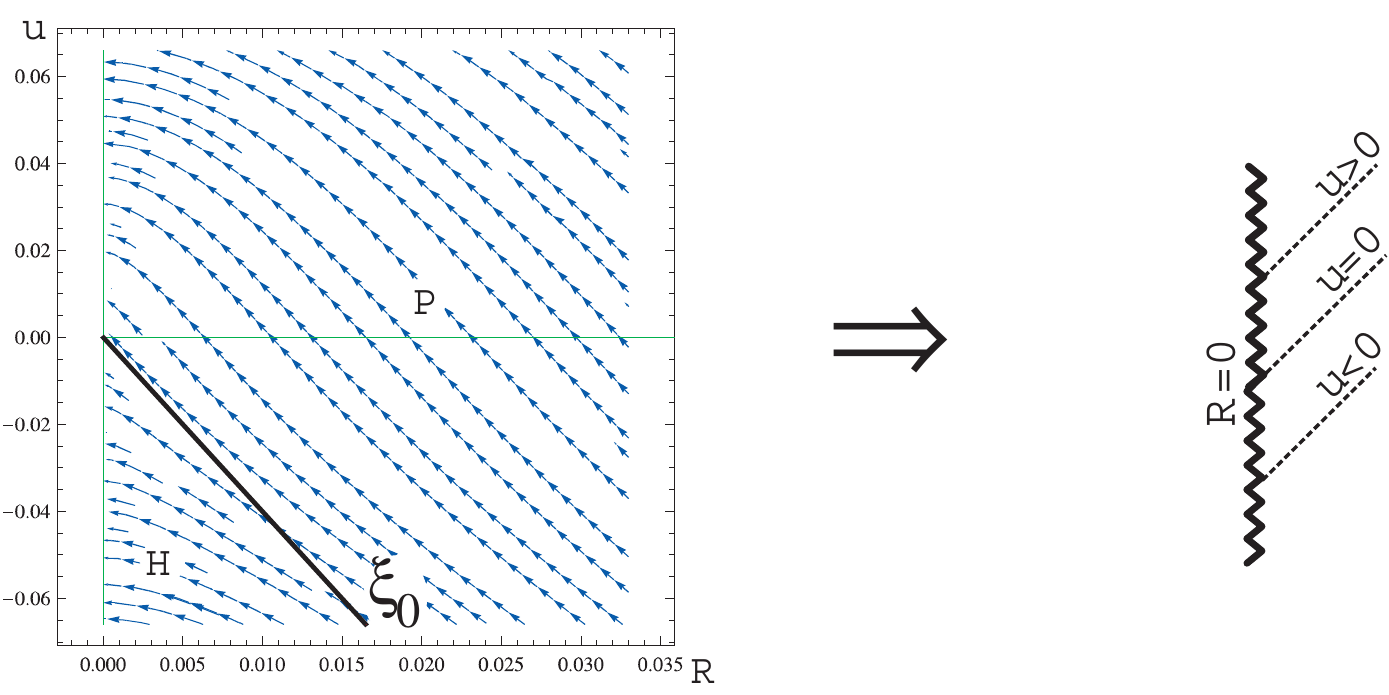}
\caption{\label{time} In the left of this figure we sketch the behaviour of the second family of radial null geodesics
in the $\varepsilon=-1$ case near $u=R=0$ when $\bar{m},_u(0,0)=0$, $\Delta_n>0$ with $n$ even and $\bar{m},_R(0,0)<1/2$. We also show by means of a straight line the regular finite critical direction $\xi_0$ of the curve starting at $u=R=0$ that separates an hyperbolic sector (H) from the parabolic sector (P) for $R\geq0$. In the right we show the corresponding Penrose diagram around $u=R=0$ with its timelike singularity.}
\end{figure}

The theory of the qualitative behaviour of dynamical systems tell us that we only need to know whether $n$ is even or odd and the sign of $\Delta_n$ to tell the qualitative behaviour of the geodesic curves around the critical point which can now be a \textit{saddle node}, a \textit{topological saddle} or a \textit{topological node}. In fact, the \textit{saddle node} behaviour is the only possibility that we have not treated yet. While all saddle nodes are the union of one parabolic and two hyperbolic sectors \cite{Andronov}, we still have to study the geodesics only in $\mathcal U$, what implies $R\geq0$. There appear four different possibilities that we show in the self-explanatory figures \ref{space}, \ref{spacelight}, \ref{timelight} and \ref{time} for the case $\varepsilon=-1$.


\begin{figure}
\includegraphics[scale=0.7]{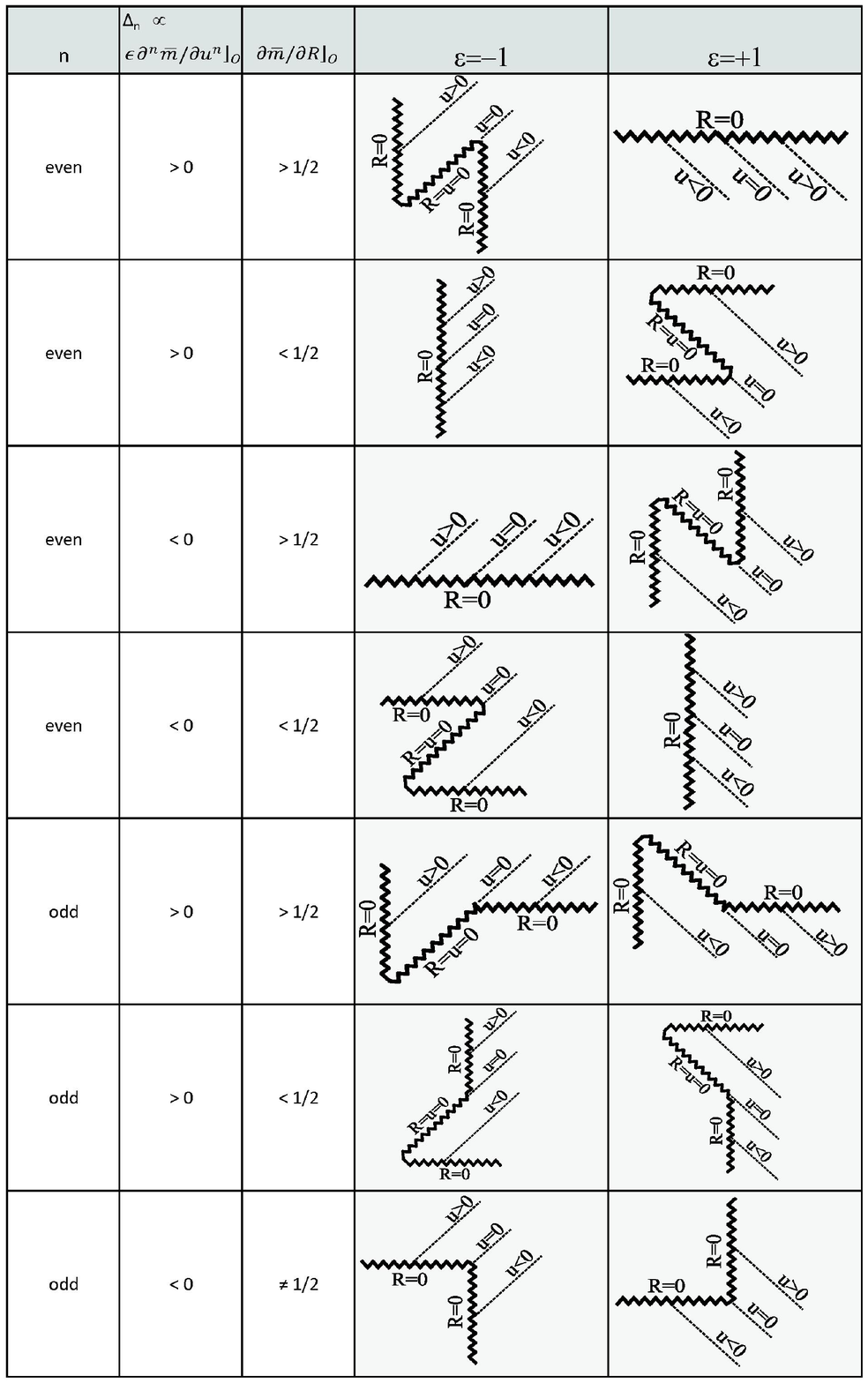}
\caption{\label{semihyper} Characterization of the singularity when $\bar{m}(u=0,R=0)=0$, $\bar{m},_u(u=0,R=0)=0$ and $\bar{m},_R(0,0)\neq 1/2$.}
\end{figure}

We have collected all the possibilities for $\bar{m},_u(u=0,R=0)=0$ and $\bar{m},_R(0,0)\neq 1/2$ in figure \ref{semihyper}.

Finally, let us analyze the case with $\bar{m},_u(0,0)=0$ and $\bar{m},_R(u=0,R=0)= 1/2$. In this case the critical point will be \textit{nilpotent} \cite{DLLA}. Again we refer the reader to \cite{Andronov}\cite{DLLA} for the theory of the qualitative behaviour of these points and to appendix B for a summary. In order to apply the theory of nilpotent points we must demand the existence of, first, a natural number $n\geq 2$ defined by $n=Max\{i,k\}$, where $i$ is the lowest value of $j$ such that $\lim_{(u\rightarrow 0,R\rightarrow 0)} \partial^j m/\partial u^j (u,R)\neq 0$ and $k=l+1$, where $l$ is the lowest value such that $\lim_{(u\rightarrow 0,R\rightarrow 0)} \partial^l (\partial m/\partial R) /\partial u^l \neq 0$ and, second, $C^n$ extensions\footnote{Private communication with the authors of \cite{DLLA},  where  $C^\infty$ is assumed independently of the value of the finite $n$.} $\bar{m}$ and $\bar{\beta}$.


In this case we simply define $x=u/2$, $y=R$, $t=\kappa$ in order to rewrite the system (\ref{system}) in the \textit{normal form}
\begin{equation*}
\left\{
\begin{array}{l}
 \frac{dx}{dt}= y+ P_2(x,y)\\
\frac{dy}{dt}= Q_2(x,y),
\end{array}
\right.
\end{equation*}
where
\begin{eqnarray}
  P_2(x,y)&\equiv& \tilde{P}_2(u(x),R(y)), \ \ \ \ \  \tilde{P}_2(u,R)= R (e^{-2\beta(u,R)}-1),\nonumber\\
  Q_2(x,y)&\equiv& \tilde{Q}_2(u(x),R(y)), \ \ \ \ \  \tilde{Q}_2(u,R)= \varepsilon (R-2 m(u,R))\nonumber
\end{eqnarray}
and they satisfy $P_2(0,0)=Q_2(0,0)=P_2,_x(0,0)=Q_2,_x(0,0)=P_2,_y(0,0)=Q_2,_y(0,0)=0$.
The solution of the equation $y+ P_2(x,y)=0$ which, by the implicit function theorem in general takes the form $y=\varphi(x)$, in this case is simply $y=0$. Following the general procedure, If we define the function $\psi(x)=Q_2(x,\varphi(x))$ then the assumed degree of differentiability guarantees that its truncated series expansion will have the form
\begin{equation}
\psi(x)= a_k x^k+...
\end{equation}
where $k\geq 2$ is assumed to be finite and, considering the Taylor polynomial for $\bar m$, one gets
\begin{equation}
a_k= -\varepsilon\frac{ 2^{k+1}}{k!} \frac{\partial^k \bar{m}}{\partial u^k}(0,0).\label{ak}
\end{equation}
On the other hand, if we define $\sigma(x)=P_2,_x(x,\varphi(x))+Q_2,_y(x,\varphi(x))$ then either its truncated series expansion can be written as
\begin{equation}
\sigma(x)=b_n x^n+...
\end{equation}
where
\begin{equation}
b_n= - \varepsilon \frac{ 2^{n+1}}{n!} \frac{\partial^n}{\partial u^n} \left( \frac{\partial \bar{m}}{\partial R}  \right )(0,0) \label{bn}
\end{equation}
if $b_n\neq 0$ or, alternatively, $\sigma(x)\equiv 0$ in which case $b_n=0\ \forall n$.

\begin{figure}
\includegraphics[scale=0.7]{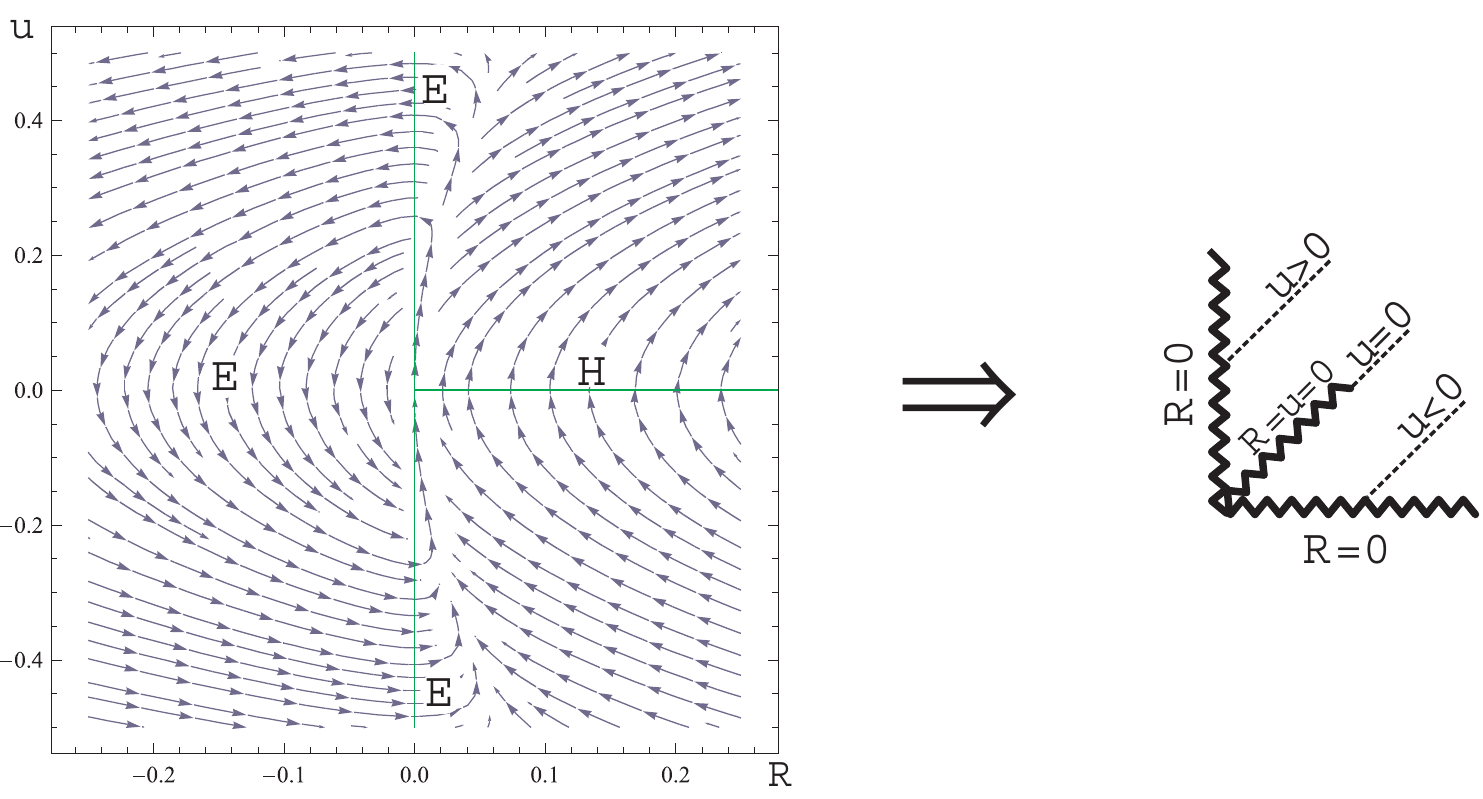}
\caption{\label{elliptic2null} In the left of this figure we sketch the behaviour of the second family of radial null geodesics
in the $\varepsilon=-1$ case near $u=R=0$ when $\bar{m},_u(0,0)=0$, $\bar{m},_R(0,0)=1/2$, $a_k<0$ with $k$ odd, $b_n>0$ with $n$ odd and either $n<(k-1)/2$ or $n=(k-1)/2$ and $\lambda\equiv b_n^2+2(k+1) a_k\geq0$. Note that we also show the behaviour for negative $R$. The fact that the hyperbolic sector (H) and the elliptic sector (E) that constitute the \textit{elliptic region} appear for $R\geq0$ and for all $u$ in $\mathcal U$ induces the unusual light-like singularity that we have drawn in the right. Specifically, all the radial null geodesics (of the second family) that come from the $R=0$-singularity for $0<u<-\delta u$ must reach in their future $R=u=0$, what implies that there is a future lightlike singularity at $R=u=0$ (see figure \ref{trans}(vii)). On the other hand, all the radial null geodesics that reach $(R=0, 0>u>\delta u)$ come from $R=u=0$, what implies that there is also a past lightlike singularity at $R=u=0$ (see figure \ref{trans}(vi)).}
\end{figure}

The theory of the qualitative behaviour of dynamic systems tell us that we only need to know whether $n$ is even or odd and the sign of $a_k$ and $b_n$ to tell the qualitative behaviour of the null curves around the critical point which can now be a \textit{saddle node}, a \textit{topological saddle}, a \textit{topological node}, a \textit{cusp}, a \textit{focus-center} or an \textit{elliptic region}. In fact, only the equilibrium state with \textit{elliptic region} behaviour provide us with a behaviour for the singularity that we have not encountered so far. We show this possibility in figure \ref{elliptic2null}.

\begin{figure}
\includegraphics[scale=0.7]{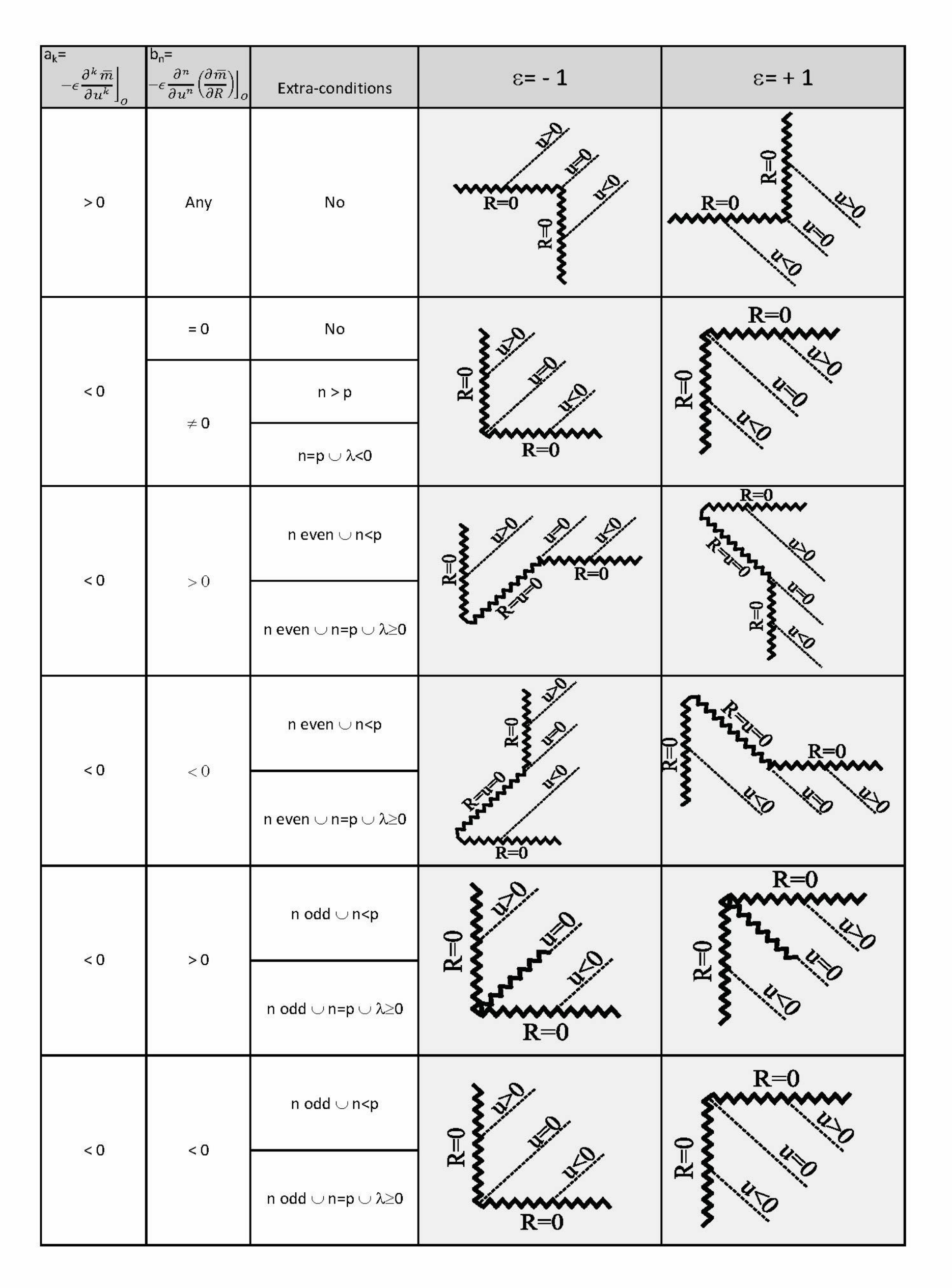}
\caption{\label{nilpotentodd} Characterization of the singularity when $\bar{m}(u=0,R=0)=0$, $\bar{m},_u(u=0,R=0)=0$, $\bar{m},_R(0,0)= 1/2$ and $k$ is odd. In the \textit{extra-conditions} we have use $p\equiv(k-1)/2$ and $\lambda\equiv b_n^2+2(k+1) a_k$. The reader can check (see appendix B or \cite{Andronov,DLLA}) that the first row corresponds to a topological saddle, the second to a focus-center, the third and fourth to a topological node and the fifth and sixth to an elliptic region. Note that the topological node and the elliptic region have two possibilities depending on the sectors appearing in $R>0$, what depends on the sign of $b_n$.}
\end{figure}

\begin{figure}
\includegraphics[scale=0.7]{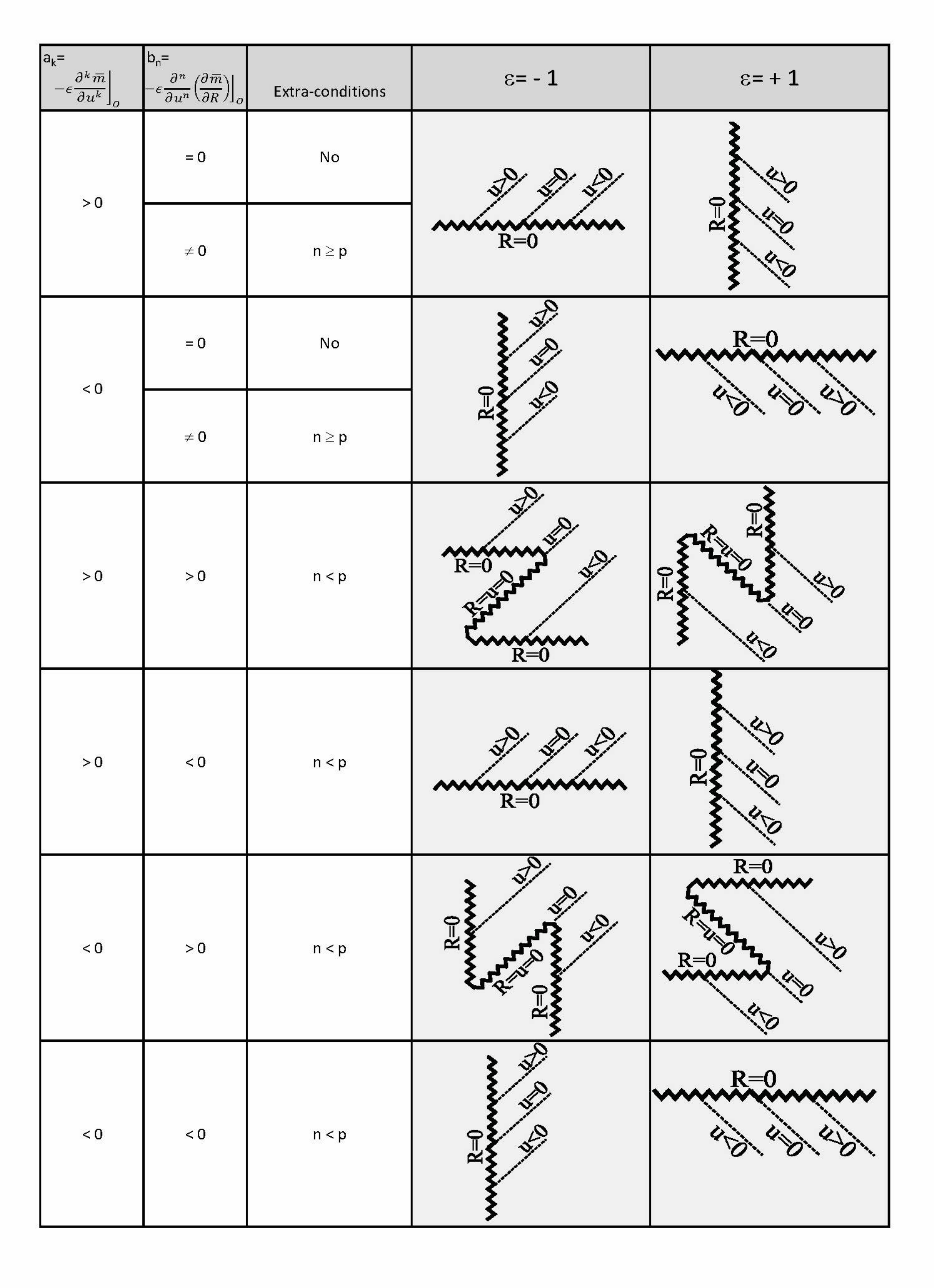}
\caption{\label{nilpotenteven} Characterization of the singularity when $\bar{m}(u=0,R=0)=0$, $\bar{m},_u(u=0,R=0)=0$, $\bar{m},_R(0,0)= 1/2$ and $k$ is even. In the \textit{extra-conditions} we have use $p\equiv k/2$. The reader can check (see appendix B or \cite{Andronov,DLLA}) that the first and second rows correspond to cusps and the rest to saddle nodes. Note that both the cusp and the saddle node have different possibilities depending on the sectors appearing in $R>0$, what now depends on the sign of $a_k$ (and also of $b_n$ in the saddle node case).}
\end{figure}

We have collected all the possibilities for $\bar{m},_u(u=0,R=0)=0$ and $\bar{m},_R(0,0)= 1/2$ in figures \ref{nilpotentodd} and \ref{nilpotenteven} with supplementary details in their corresponding captions.

The cases analyzed so far cover all the possibilities whenever $u=R=0$ is an isolated critical point.

\section{Characterization when $\bar{m}(-\delta u<u<\delta u,R=0)=0$}\label{NICP}
\begin{figure}
\includegraphics[scale=0.7]{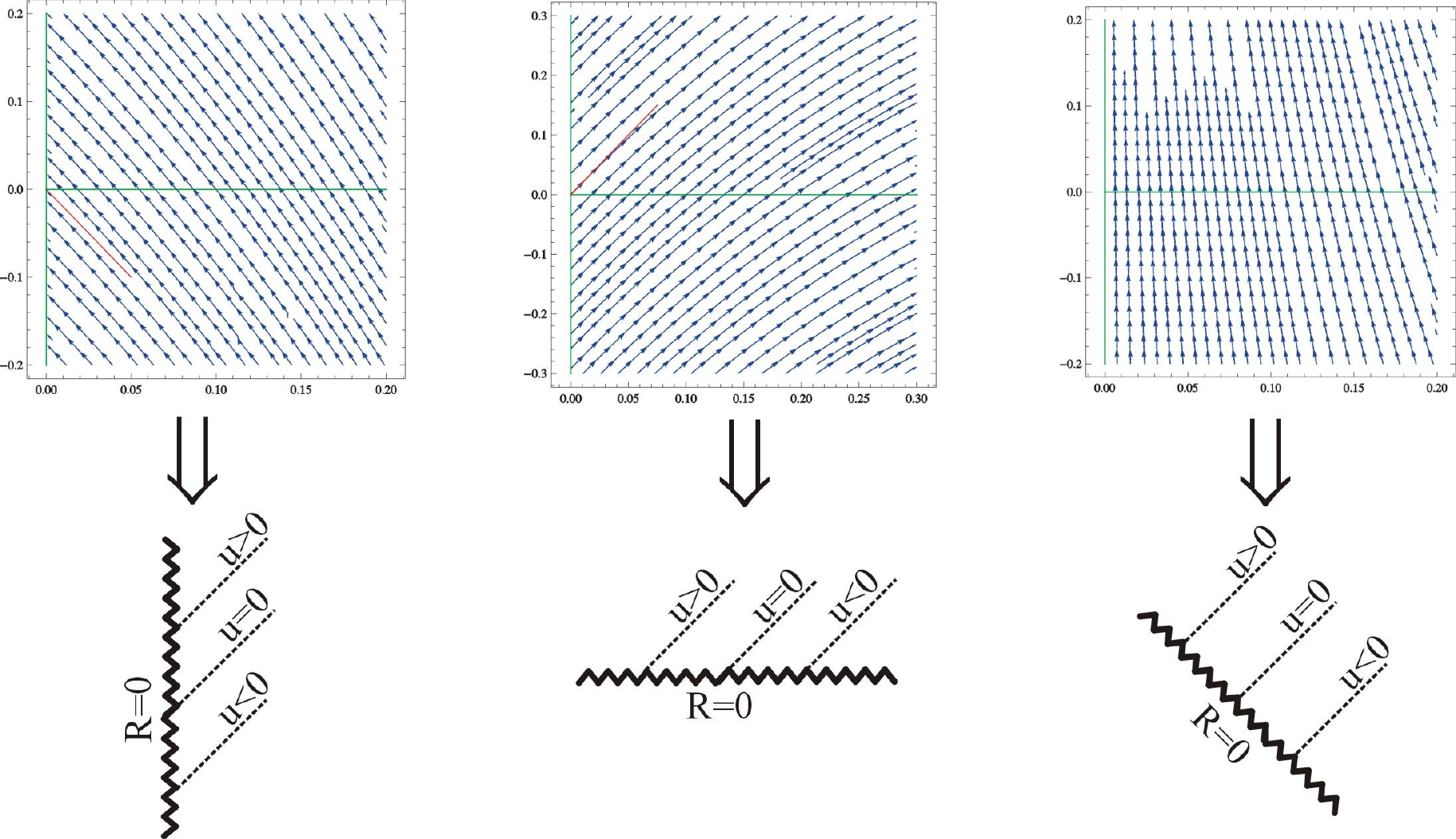}
\caption{\label{noniso} We have sketched above the $u-R$ graphics for the three different cases when there is a non-isolated critical point and $\varepsilon=-1$. The order is, from the left to the right, $\bar{m},_R(u,R=0)<1/2$, $\bar{m},_R(u,R=0)>1/2$ and $\bar{m},_R(u,R=0)=1/2$. Below these graphics we have drawn the corresponding sketched Penrose diagrams.}
\end{figure}

In this case the curves $f^R(u,R)\equiv\varepsilon [R-2 \bar{m}(u,R)]=0$ and $f^u(u,R)\equiv 2 R e^{-2\bar{\beta}}=0$ must cross in an interval made up of points which are non-isolated critical points for the system of differential equations (\ref{system}).
Assuming the existence of the $C^1$ extensions $\bar m$ and $\bar \beta$, the equation for the radial null geodesics (\ref{system}) can be easily written on the interval, up to first order, as
\begin{equation}
\zeta(u)\equiv\frac{dR}{du}(u,R=0)=\frac{\varepsilon (1-2 \bar{m},_R(u,0))}{2}, \label{zeta}
\end{equation}
where we have used the fact that $\bar{m}(-\delta u<u<\delta u,R=0)=0$ implies that $\bar{m},_u=0$ in the interval.
Let us consider the characterization of the singularity for the case $\varepsilon=-1$ (the case $\varepsilon=+1$ can be easily obtained later on as its time-reversal). Then the u=constant radial null geodesics are outgoing while
\begin{itemize}
\item If $\bar{m},_R(u,R=0)<1/2$ in $\bar{\mathcal U}$ the second family of radial null geodesics satisfy $\zeta<0$ and they are ingoing. Therefore, the singularity there\footnote{In case there is a \textit{singularity}, since this case includes the \emph{regular} one (\ref{condisreg}).} is time-like.
\item If $\bar{m},_R(u,R=0)>1/2$ in $\bar{\mathcal U}$ the second family of radial null geodesics satisfy $\zeta>0$ and they are outgoing. Therefore, the singularity there is space-like.
\item If $\bar{m},_R(u,R=0)=1/2$ in $\bar{\mathcal U}$ then $\zeta=0$.
 In this way, the trajectories of the second family of radial null geodesics tend to be parallel to the $R=0$ interval the closer they are to the singularity and, thus, the singularity is light-like.
\end{itemize}
We have sketched the three situations in figure \ref{noniso} for $\varepsilon=-1$. (It is not necessary to sketch the case $\varepsilon=+1$ since it produces the time reversal diagrams for every case).

\section{Concluding remarks and some applications}

Let us assume that the reader finds a specific singular model
and wants to check the causal character of its $R=0$-singularity
by means of the results provided in this article. In order to clarify the applicability of our approach and to make the reader's task easier, we would now like to summarize the assumptions that have been made along the article and the path that the reader should follow.
On the one hand, we have assumed that
\begin{itemize}
\item[---] We are working with a time-orientable spherically symmetric spacetime possessing a local chart endowed with coordinates $\{u,R,\theta,\varphi\}$ (see section \ref{GSSM}). In this way, we assume that the invariant areal radius $R$ can be used as a coordinate in our local chart.
\item[---] The metric, written as in (\ref{mI}), depends on functions $\beta$ and $\chi$ which are at least $C^{2-}$ in the local chart (this is a minimum requirement in order to guarantee the existence and uniqueness of null geodesics. Note however that the degree of differentiability is usually required to be higher). $\beta$ is also assumed to be bounded.
\item[---] The local study is carried out for finite values of $u$ and, specifically, we have chosen to work around $u=0$ (if necessary, it suffices a simple coordinate change $u=\bar u+$constant). We assume that, at least, (future or past directed) radial null geodesics of the $\mathcal{F}_1$ family ($u=$constant) reach $R=0$ in an interval around $u=0$ ($-\delta u<u<\delta u$).
\item[---] The spacetime has a singular boundary that is (in the unphysical spacetime), at least, piecewise $C^1$.
\item[---] In the cases where an extension ($\bar m$) of the invariant $m$ is required, then we have not considered working around non-isolated critical points of (\ref{system}) that are accumulation points of the set $\bar{ m}(R=0,u)=0$.
\end{itemize}

On the other hand, the reader should be aware that, for every case, our use of the qualitative theory of dynamic systems implies that the complete path could only be followed
if some extra-requirements on the differentiability of the functions $m$ and $\beta$
are satisfied.
Let us then summarize the aforesaid path along with every differentiability requirements:
\begin{itemize}

\item If $\lim_{(u\rightarrow u_0,R\rightarrow 0)} m(u,R)\neq 0$ ($\forall u_0 \in \mathcal U$, i.e., for an interval of $u$'s) there is not assumption required here\footnote{Except for the obvious minimum degree of differentiability ($C^{2-}$) for $m$ and $\beta$ \emph{in} $\mathcal U$.}. The limit can be either finite or infinite and the characterization depends only on whether it is positive (spacelike singularity) or negative (timelike singularity) (final corollary in section \ref{csmneq0}).

    \subitem Note that if $\lim_{(u\rightarrow u_0,R\rightarrow 0)} m(u,R)$ \emph{does not exist}
then we can still evaluate $m$ along the radial null geodesics of the second family: $\lim_{\kappa\rightarrow \kappa_0} m(u,R)$ (see section \ref{csmneq0}). If it is not zero, no matter if it is finite or infinite, the characterization depends only on whether it is positive (piecewise spacelike singularity) or negative (piecewise timelike singularity). On the other hand, if only the directional limit exists and it is zero the method is not conclusive.

\item If $\lim_{(u\rightarrow u_0,R\rightarrow 0)} m(u,R)=0$ ($\forall u_0 \in \mathcal U$, i.e., for an interval of $u$'s) then we demand the existence of $C^1$ extensions for $m$ and $\beta$. If $\lim_{(u\rightarrow u_0,R\rightarrow 0)} m,_R (u,R)$ is less than $1/2$ in $\mathcal U$ then there is a timelike singularity, if the limit equals $1/2$ in $\mathcal U$ then there is a lightlike singularity and if it is greater than $1/2$ then there is a spacelike singularity (section \ref{NICP}).

\item If $\lim_{(u\rightarrow 0,R\rightarrow 0)} m(u,R)=0$ and $\lim_{(u\rightarrow 0,R\rightarrow 0)} m,_u (u,R)\neq 0$ then we demand the existence of a $C^1$ extension for $m$ and $\beta$ (section \ref{m0hyp}). In this case the results can be found in figure \ref{hyper}.

\item If $\lim_{(u\rightarrow 0,R\rightarrow 0)} m(u,R)=0$ (and only for $u\rightarrow 0$ ), $\lim_{(u\rightarrow 0,R\rightarrow 0)} m,_u (u,R)= 0$ and $\lim_{(u\rightarrow 0,R\rightarrow 0)} m,_R (u,R)\neq 1/2$ then we demand, first, the existence of a natural number $n\geq 2$ such that $\lim_{(u\rightarrow 0,R\rightarrow 0)} \partial^n m/\partial u^n (u,R)\neq 0$, while $\lim_{(u\rightarrow 0,R\rightarrow 0)} \partial^i m/\partial u^i (u,R)= 0$ for $i=1,...,n-1$, and, second, of $C^n$ extensions $\bar{m}$ and $\bar{\beta}$ (section \ref{m0nonhyp}). In this case the results can be found in figure \ref{semihyper}.

\item If $\lim_{(u\rightarrow 0,R\rightarrow 0)} m(u,R)=0$ (and only for $u\rightarrow 0$), $\lim_{(u\rightarrow 0,R\rightarrow 0)} m,_u (u,R)= 0$ and $\lim_{(u\rightarrow 0,R\rightarrow 0)} m,_R (u,R)= 1/2$ then we demand the existence of, first, a natural number $n\geq 2$ defined by $n=Max\{i,k\}$, where $i$ is the lowest value of $j$ such that $\lim_{(u\rightarrow 0,R\rightarrow 0)} \partial^j m/\partial u^j (u,R)\neq 0$ and $k=l+1$, where $l$ is the lowest value such that $\lim_{(u\rightarrow 0,R\rightarrow 0)} \partial^l (\partial m/\partial R) /\partial u^l \neq 0$ and, second, $C^n$ extensions $\bar{m}$ and $\bar{\beta}$ (section \ref{m0nonhyp}). In this case the results can be found in figures \ref{nilpotentodd} and \ref{nilpotenteven}.

\end{itemize}

\begin{figure}
\includegraphics[scale=0.8]{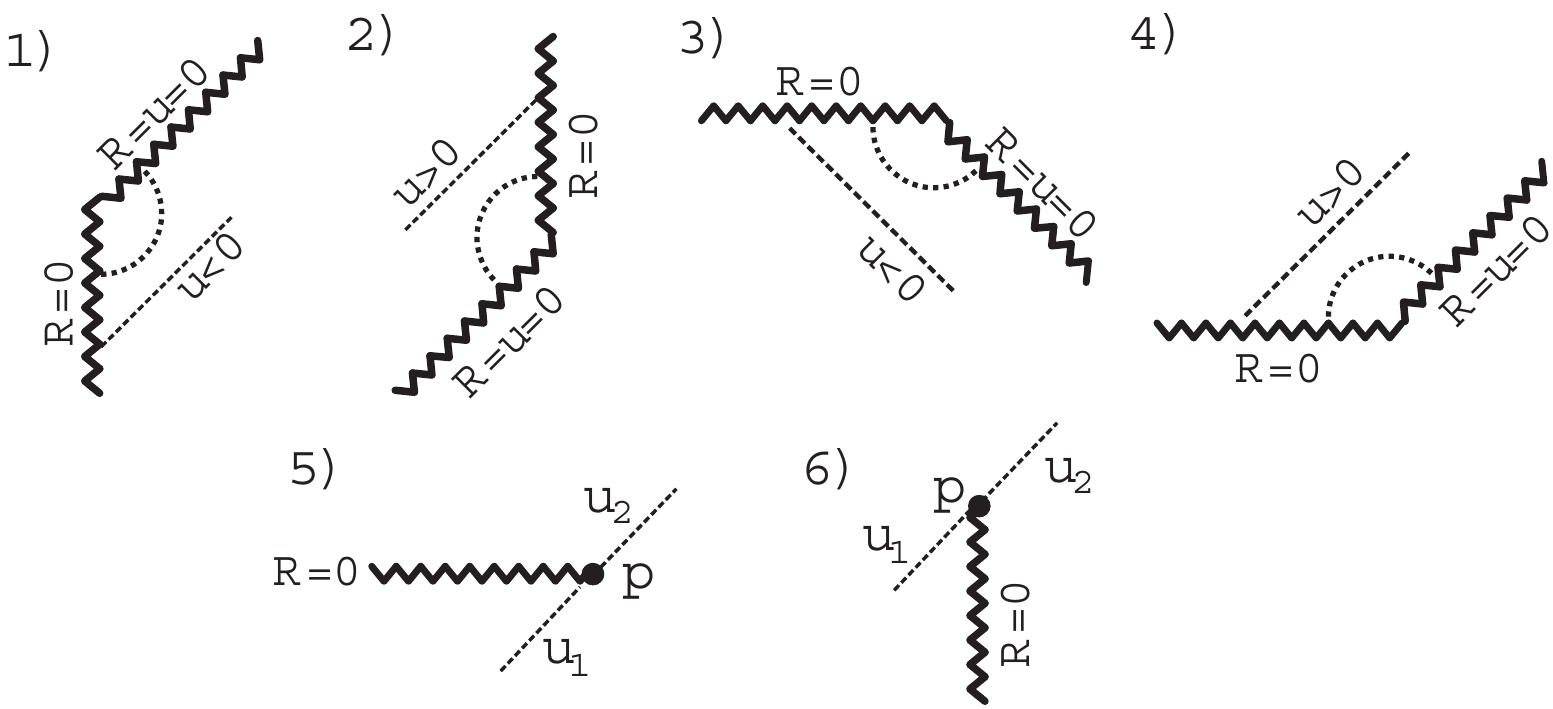}
\caption{The forbidden possibilities if one works with a single local chart and assuming the existence of the required $C^n$ extension. From 1) to 4) they are particular transitions from timelike to lightlike singularities and from spacelike to lightlike singularities forbidden by the requirement of the degree of differentiability for the extension. Specifically, the forbidden possibilities are those that in our sketches are represented forming an angle of 135 degrees in the spacetime. From 5) to 6) we show the \textit{pointed} singularities that can not be described around $p$ with a single local chart in our coordinates. We explicit a possibility showing that the $u=$constant geodesics cannot be always ingoing or outgoing as required (in both cases $u_1$ is ingoing while $u_2$ is outgoing).
\label{135s}}
\end{figure}

Assuming that we use a single local chart and that the required $C^n$ extensions exist for every case, an interesting question appears: Are there any forbidden possibilities for the causal behaviour of the singular boundary? If one inspects the figures in the article it is easy to see that, indeed, some possibilities do not appear. Specifically, there are four possibilities which do not appear due to the degree of differentiability required. We have collected them under the name of ``135º singularities" (see figure \ref{135s}(1-4)). This is expected from the qualitative theory of dynamic systems since the theorems involved only allow for specific combinations of sectors. On the other hand, the use of a local chart does no allow us to work \emph{around} transition points that can be reached by two radial null geodesics and left by another two
(\textit{pointed} singularities, see figure \ref{135s}(5 and 6)) since, as explained in subsection \ref{nullgeodesics}, the $u=$constant radial null geodesics must always be outgoing or ingoing in a single local chart.

\subsection{The role of the scalar invariant $m$ in the causal characterization}

As we have seen, despite the function $\beta$ is involved in the behaviour of the radial null geodesics (\ref{system}), the causal characterization of the zero-areal-radius singularity around $u=R=0$ for all the different possibilities depends in quantities ($\Delta$, $T$, det$A$, $\Delta_n$, $a_k$, $b_n$) that are just obtained from the knowledge of the lowest-order non-zero derivatives of $m(u,R)$ as $u$ and $R$ tend to zero. Then we can state the following result:
\begin{propos}
Provided the listed assumptions are satisfied the knowledge of the scalar invariant $m$ is all that is required in order to obtain the sketched causal characterization of the $R=0$ singularity in a spherically symmetric space-time.
\end{propos}

We can refine this result taking into account that the knowledge of $m$
is not a \textit{necessary}, but a \emph{sufficient} condition in order to characterize the singularity in this approach. By inspecting the quantities defining the character of the singularity we see that the knowledge of the lowest non-zero $\lim_{ u\rightarrow 0, R\rightarrow 0} \partial^k m/\partial u^k $ ($k>0$) and $\lim_{ u\rightarrow 0, R\rightarrow 0} \partial^n m,_R/\partial u^n (0,0)$ ($n>0$) suffices to characterize the singularity. Then, assuming the existence of the extension for $m$ with its corresponding degree of differentiability and using Taylor's theorem up to that degree 
we see that a rough knowledge of $\bar{m}(u,R=0)$ and $\bar{m},_R(u,R=0)$ (note that they are evaluated just at $R=0$) suffices to characterize any $R=0$ singularity in $\mathcal U$.

\subsection{Lightlike singularities and shell-focussing nakedness}

When dealing with naked singularities that form due to the gravitational collapse, one usually looks for the generation of timelike or lightlike singularities. Provided that our assumptions are satisfied, the timelike case is straightforward; it requires either $lim_{R\rightarrow 0} \ m<0$ in an interval of $u$'s in $\mathcal U$\footnote{Think, for instance, of the timelike singularity in the Reissner-Nordstr\"{o}m solution, where a negative limit is justified due to the presence of electrical charge.} (section \ref{csmneq0}) or $lim_{R\rightarrow 0} m=0$ (if the extra condition $lim_{R\rightarrow 0} m,_R(u,R)<1/2$ is satisfied for some proper interval of $u$'s in $\mathcal U$) (section \ref{NICP}).
The lightlike case, however, requires a careful study to which we will devote this subsection. For example, assuming the existence of the corresponding extension for $m$ and by inspecting the hyperbolic and the semi-hyperbolic cases (figures \ref{hyper} and \ref{semihyper}, respectively) we arrive to the
\begin{propos}
Assuming the existence of, first, a natural number $n\geq 1$ such that $\lim_{(u\rightarrow 0,R\rightarrow 0)} \partial^n m/\partial u^n (u,R)\neq 0$ (while, in case $n\neq 1$, $\lim_{(u\rightarrow 0,R\rightarrow 0)} \partial^i m/\partial u^i (u,R)= 0$ for $i=1,...,n-1$) and, second, of $C^n$ extensions $\bar{m}$ and $\bar{\beta}$,
%
if a $R=0$ singularity satisfies
$\bar{m}(0,0)=0$,
\begin{eqnarray}
\varepsilon^n  \frac{\partial^n \bar{m}}{\partial u^n}(0,0) \left[ 1-2 \bar{m},_R(0,0)\right]^{n+1} & > & 0, \\
\left[ 1-2 \bar{m},_R (0,0)\right]^{2} -16 \varepsilon \bar{m},_u(0,0)&\geq& 0,
\end{eqnarray}
then it has a light-like singularity at $R=u=0$.

Furthermore, if $\varepsilon (1-2 \bar{m},_R(0,0))>0$ there is a \emph{past} lightlike singularity, whereas if $\varepsilon (1-2 \bar{m},_R(0,0))<0$ there is a \emph{future} lightlike singularity.
\end{propos}

As a particular application let us consider the case of Vaidya's radiating solution. Since in this solution $m$ only depends on $u$ we will have
\begin{coro}
Any Vaidya metric admitting, first, a natural number $n\geq 1$ such that $\lim_{u\rightarrow 0} d^n m(u)/d u^n \neq 0$ (while, in case $n\neq 1$, $\lim_{u\rightarrow 0} d^i m/d u^i = 0$ for $i=1,...,n-1$) and, second, a $C^n$ extension $\bar{m}$,
%
will develop a light-like singularity at $R=u=0$ if
\begin{eqnarray*}
\bar{m}(u=0)=0,\\
\varepsilon \frac{d\bar{m}}{du}(u=0)\leq \frac{1}{16},\\
\varepsilon^n \frac{d^n \bar{m}}{d u^n}(u=0) >0.
\end{eqnarray*}
\end{coro}

This corollary generalizes previous results on Vaidya's metric with $\varepsilon=+1$ for the linear \cite{HisWiEar}\cite{Papapetrou} and the non-linear \cite{Kuroda} cases.

Among the lightlike singularities some of them are \textit{persistent naked singularities}. By ``persistent" we denote a naked singularity such that a \emph{whole family} of future-directed lightlike radial null geodesics emerges from it.
It has been pointed out that these naked singularities can appear as a consequence of shell-focusing \cite{HisWiEar}\cite{E&S}\cite{Chris}\cite{O&P}\cite{Papapetrou}\cite{C&S} during the collapse and that they are, therefore, relevant to the cosmic censorship conjecture. If we want to avoid that the 2-spheres close enough to the $R=0$-singularity become past-trapped closed surfaces (usually only \emph{future}-trapped 2-surfaces are considered as possible byproducts of collapse) we should also demand $\varepsilon=+1$ (see subsection \ref{nullgeodesics}). By filtering the lightlike singularities with these properties in the above proposition we arrive to the following
\begin{propos}\label{pllns}
A general spherically symmetric spacetime with metric (\ref{mI}) admitting, first, a natural number $n\geq 1$ such that $\lim_{(u\rightarrow 0,R\rightarrow 0)} \partial^n m/\partial u^n (u,R)\neq 0$ (while, in case $n\neq 1$, $\lim_{(u\rightarrow 0,R\rightarrow 0)} \partial^i m/\partial u^i (u,R)= 0$ for $i=1,...,n-1$) and, second, $C^n$ extensions $\bar{m}$ and $\bar{\beta}$,
will develop a persistent lightlike naked singularity as a consequence of gravitational collapse if $\varepsilon=+1$, $\bar{m}(0,0)=0$, $\bar{m},_R(0,0)<1/2$,
\begin{eqnarray*}
\left[1-2 \bar{m},_R(0,0)\right]^2-16 \bar{m},_u(0,0)\geq 0,\\
\frac{\partial^n \bar{m}}{\partial u^n}(0,0)>0.
\end{eqnarray*}
\end{propos}

Note that persistent lightlike naked singularities are also possible in the especial case $\bar{m},_R(0,0)=1/2$. The reader can write explicitly the conditions for them directly from figures \ref{nilpotentodd}, \ref{nilpotenteven} and \ref{noniso}.

So far we have been working with a single local chart in $\mathcal U$ and assuming that certain extensions of $m$ exist and possess a certain degree of differentiability. Therefore, the cases in which these requirements are not fulfilled have been excluded up to now. However, in some cases, the method can be generalized to include different local charts or more general differentiability requirements.
For example, \emph{particular} models for collapsing stars with an initially regular center ($\bar{m}(u<0,R=0)=0$) in which the shell focussing eventually generates a persistent lightlike naked singularity ($u=0$) could be constructed with the usual matching techniques (see \cite{HisWiEar}\cite{Papapetrou} for particular cases), which allow the avoidance of the differentiability requirements and the limitations of the use of a single local chart.
It is only required that there is a regular spacetime modelling the initially regular interior, that there is a singular spacetime whose scalar invariant $m$ satisfies the conditions in proposition \ref{pllns} and that the two spacetimes can be matched through a non-spacelike hypersurface reaching $R=0$ at $u=0$.

In the next subsection we will also go beyond the limitations of dealing with a single local chart and with the required differentiability assumptions by working with a particular interesting novel application: Evaporating Black Holes that can develop \textit{persistent} naked singularities. With these treatments our results could be used to cover most of the spherically symmetric solutions found in the literature. Nevertheless, let us remark that one can still find particular cases where even the generalized treatment would fail. An example of this would be the model in \cite{D&L} which possesses a function $m$ that is not even well defined at $u=R=0$.

\subsection{Beyond the restrictions: Evaporating black holes developing persistent naked singularities}

\begin{figure}
\includegraphics[scale=0.7]{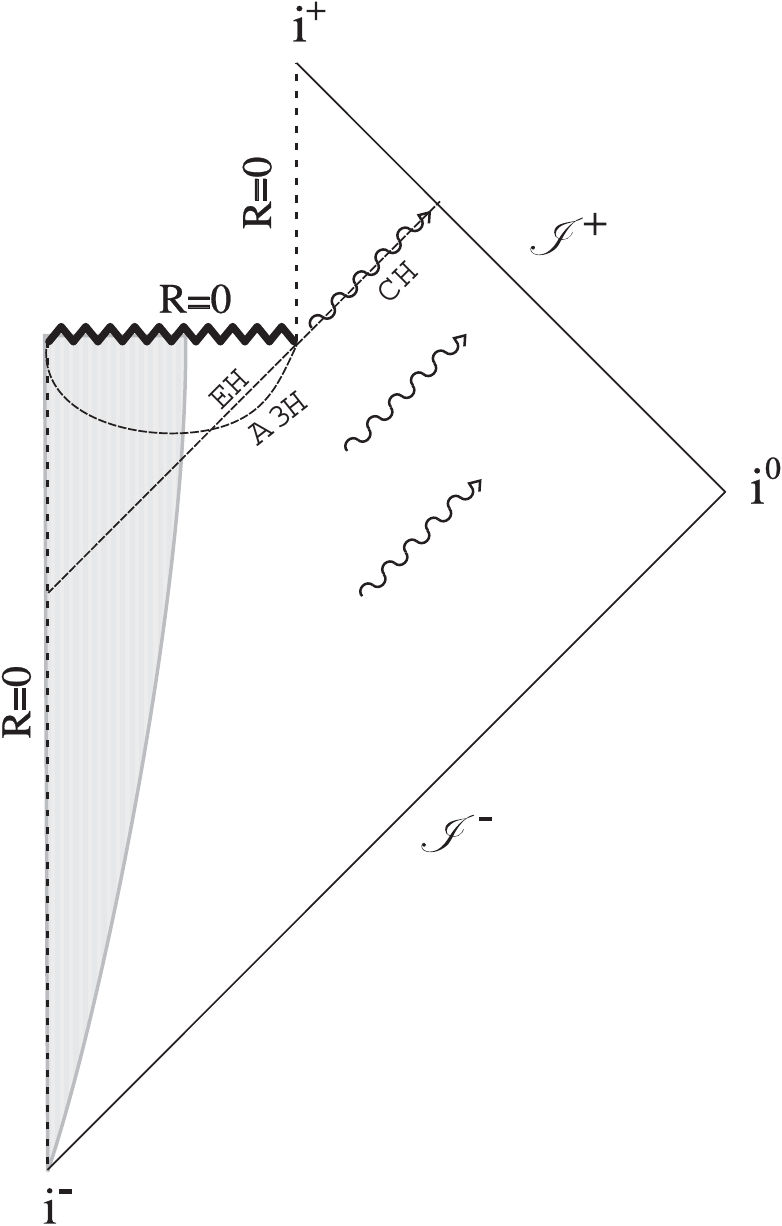}
\caption{Typical Penrose's diagram for a collapsing star generating a BH that evaporates due to the emission of Hawking radiation. The grey zone corresponds with the interior of the star. The exterior of the star
will usually contain radiation coming out from the star (the wavy arrows) and particles and radiation due to the Hawking effect.
\label{EBH}}
\end{figure}

Let us consider a future spacelike singularity that reaches a point, say at $u=0$, where $\lim_{u\rightarrow 0,R \rightarrow 0} m=0$ and then it is followed by a regular
%
$R=0$ --see (\ref{condisreg})\footnote{Note that the singularity disappears for $u>0$ provided that $\beta$ also satisfies its own regularity conditions (\ref{condisreg}) as it approaches $R=0$ for $u>0$.}-- for $u>0$. If this happens, we will say that the singularity \textit{evaporates}.
In fact, this is the usually expected behaviour for the singularity of an evaporating black hole (EBH) \cite{Wald}. (We have illustrated it with a particular \emph{complete} model in figure \ref{EBH}).
In this case there will be a future directed radial null geodesic starting at $u=R=0$ in the singularity. In this way, the singularity must be naked and this geodesic defines a Cauchy Horizon (CH) of the model. Sometimes this singularity is called an \textit{instantaneous} naked singularity \cite{L&H} since an observer crossing the CH could only detect a single flash of null radiation coming from the singularity.

\begin{figure}
\includegraphics[scale=0.7]{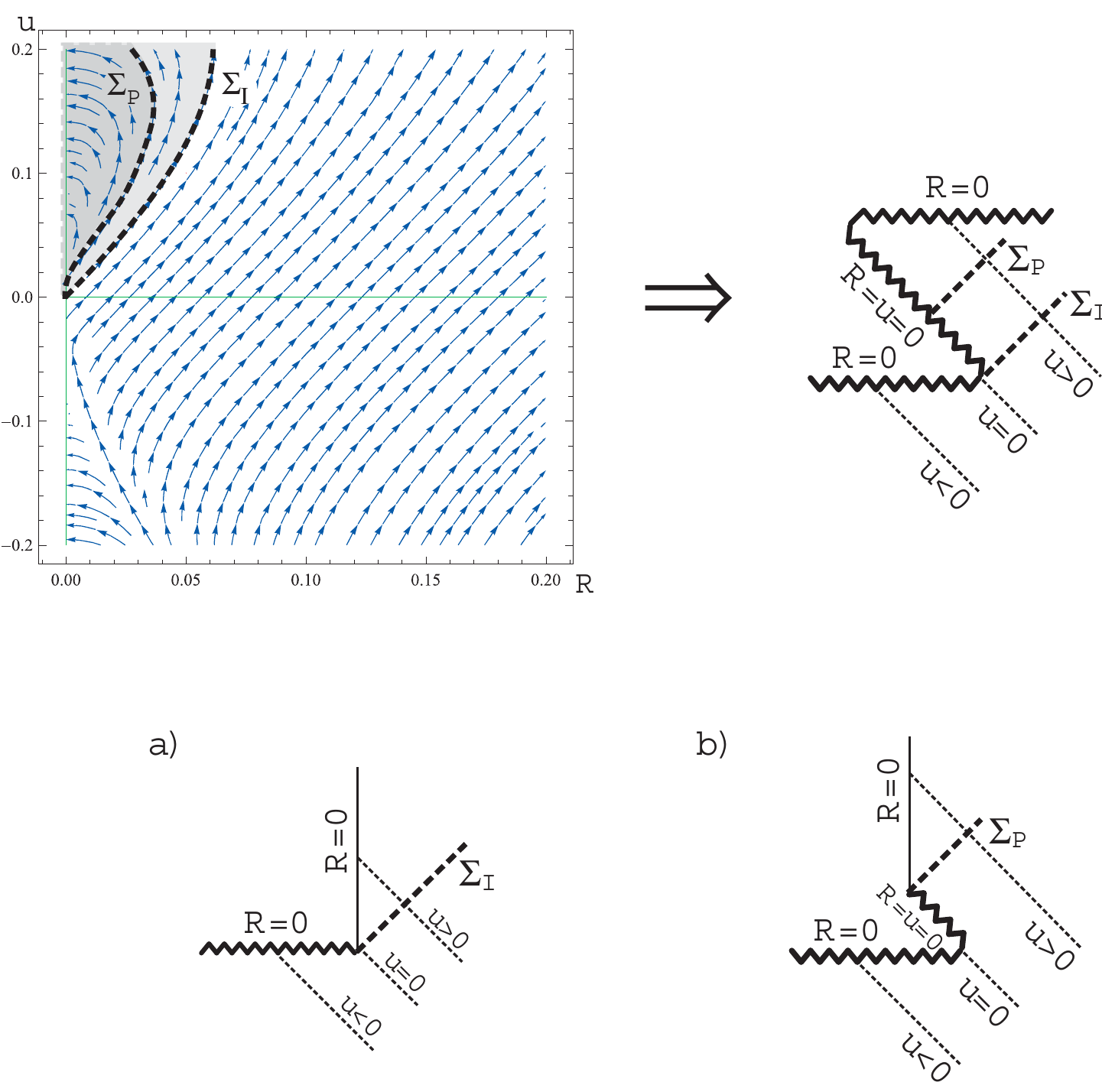}
\caption{The upper part of the graphic shows the second family of radial null geodesics for an $m$ satisfying the requirements in proposition \ref{proposPNS} together with its corresponding sketched conformal diagram. We have outlined two different lightlike hypersurfaces as our candidate for matching hypersurface: $\Sigma_I$ and $\Sigma_P$. In the first case, the only lightlike geodesic with a \textit{regular direction} $\Sigma_I$ is chosen. The grey region is replaced with a region with a regular center and the result of the matching possessing an instantaneous naked singularity is shown in a). In the second case, one of the outgoing lightlike geodesics leaving $u=R=0$ with infinite slope is chosen as the matching hypersurface $\Sigma_P$. The darker grey region is replaced with a region with a regular center and the result of the matching possessing a persistent naked singularity is shown in b).\label{match2}}
\end{figure}

\begin{figure}
\includegraphics[scale=0.7]{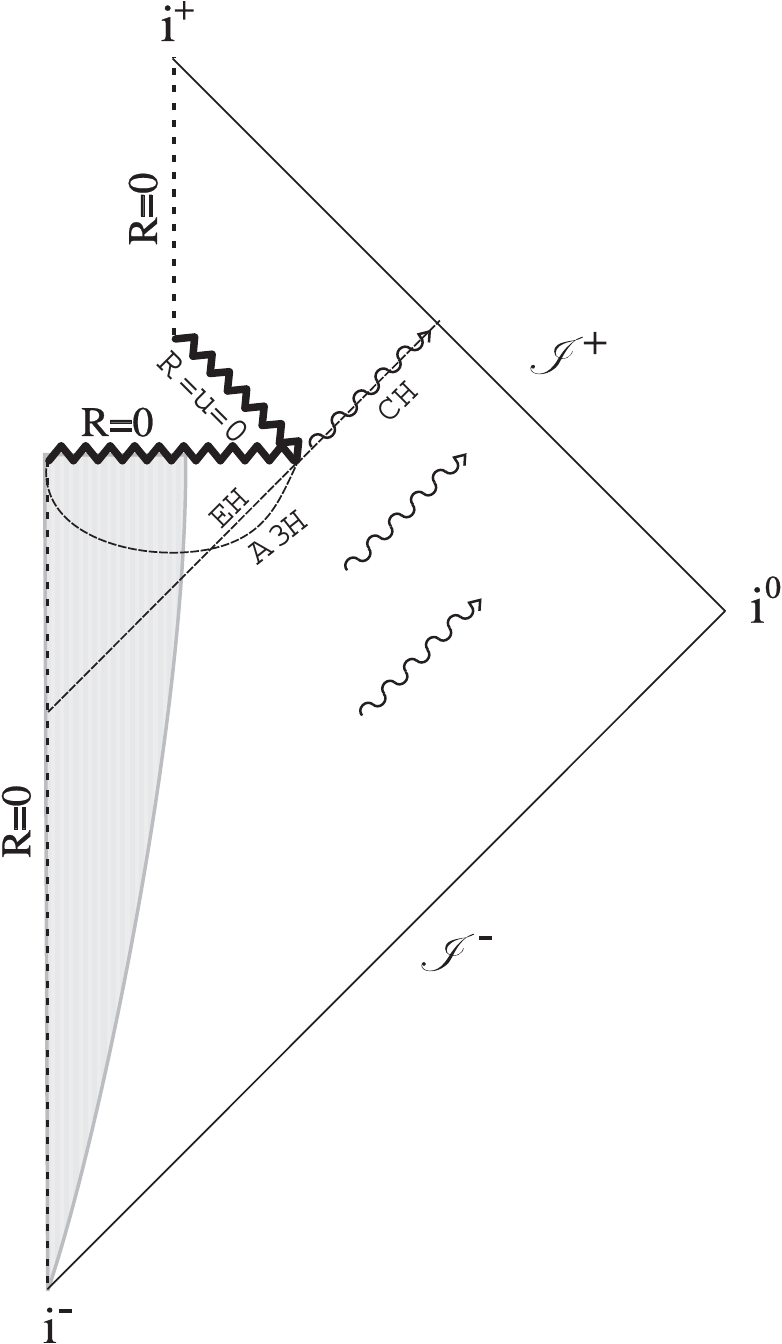}
\caption{Penrose's diagram for a collapsing star generating a BH that evaporates due to the emission of Hawking radiation. The scalar invariant $m$ satisfies the requirements in proposition \ref{proposPNS} for $u\leq 0$ and a regular center is proposed for $u>0$. The resulting diagram contains a persistent globally naked singularity: The light-like $R=u=0$ singularity. \label{PNS}}
\end{figure}

Now we would like to show that the requirement
of regularity for $u>0$
in $\mathcal U$ allows
the interesting possibility of the development of a \emph{persistent} naked singularity at $u=0$.
Specifically the results in  section \ref{m0nonhyp} imply the following
\begin{propos}\label{proposPNS}
A singular spherically symmetric spacetime with metric (\ref{mI}) admitting a natural number $n\geq 2$ such that $\lim_{(u\rightarrow 0,R\rightarrow 0)} \partial^n m/\partial u^n (u,R)\neq 0$, while $\lim_{(u\rightarrow 0,R\rightarrow 0)} \partial^i m/\partial u^i (u,R)= 0$ for $i=1,...,n-1$,
admitting $C^n$ extensions $\bar m$ and $\bar \beta$
%
and with a $R=0$ future space-like singularity will develop a persistent lightlike naked singularity at $R=u=0$ if
$\varepsilon=+1$,
\begin{eqnarray*}
\bar{m}(0,0)&=&0, \\
\bar{m},_u(0,0)&=&0, \\
\frac{\partial^n \bar{m}}{\partial u^n}(0,0)&>&0,
\end{eqnarray*}
provided that $n$ is even, and one of these two options is satisfied
\begin{itemize}
\item a)
\begin{equation}
\frac{\partial \bar{m}}{\partial R}(0,0)<\frac{1}{2}
\end{equation}
\item b)
\begin{equation}
\frac{\partial \bar{m}}{\partial R}(0,0)=\frac{1}{2}\qquad\mbox{and}\qquad \frac{\partial^i}{\partial u^i} \left( \frac{\partial \bar{m}}{\partial R} \right) (0,0)<0,
\end{equation}
where $i$ is the lowest number such that the partial derivative is non-null provided that $i<n/2$.
\end{itemize}
\end{propos}
Note that the requirement $n\geq 2$ indicates that the generation of such light-like singularities is related to the presence of a scalar invariant $m$ reaching its zero value slowly enough.
In figure \ref{match2} we have sketched a procedure for constructing a future spacelike singularity developing a persistent lightlike naked singularity and \emph{followed by a regular center} using the usual matching technique.
On the other hand, in figure \ref{PNS} we show a complete spacetime with a persistent lightlike globally naked singularity and a regular $R=0$-center for $u>0$.

\section*{Acknowledgements}
We would like to thank J.M.M. Senovilla, J. LLosa, M. S\'{a}nchez, J. Herrera, J.L. Flores and J. Llibre for helpful discussions.
We would also like to acknowledge the \textit{Ministerio de Educaci\'{o}n y Ciencia}  (FIS2007-63034) and the \textit{Generalitat de Catalunya} (grant 2009SGR-00417) for financial support.

\appendix

\section{Semi-Hyperbolic Critical Points}
\begin{figure}
\includegraphics[scale=0.6]{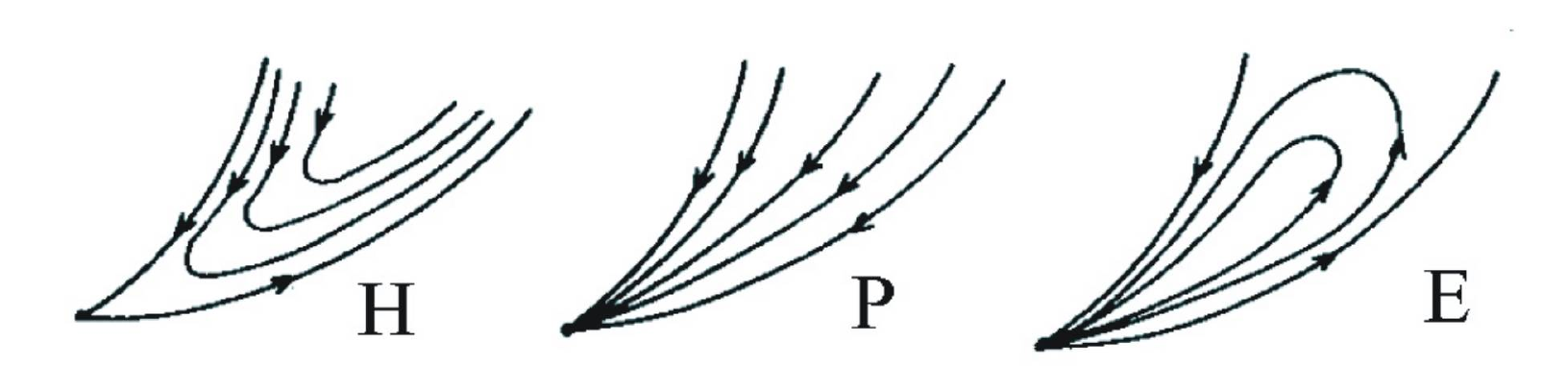}
\caption{(H) Hyperbolic sector, (P) parabolic sector and (E) elliptic sector.\label{sectors}}
\end{figure}

Let us assume that the origin is an isolated critical point of a planar system. Then, if the linearization matrix $A$ satisfies $\det A=0$ and its trace $T\neq0$ then the critical point will be \textit{semi-hyperbolic} \cite{DLLA}. The details on the theory regarding the qualitative behaviour of these points can be found in \cite{Andronov}\cite{DLLA}. We will summarize here only the main results:

The neighborhood of the origin can be divided into open regions called \textit{sectors} which can be \textit{hyperbolic}, \textit{parabolic} or \textit{elliptic} according to their topological equivalence with the figures \ref{sectors}-H, \ref{sectors}-P or \ref{sectors}-E, respectively. The trajectories which lie on the boundary of a hyperbolic sector are called \textit{separatrices}.

Given a system possessing an isolated semi-hyperbolic critical point there is a suitable linear transformation that allows us to write it as:
\begin{equation*}
\left\{
\begin{array}{l}
 \frac{dx}{dt}= P_2(x,y)\\
\frac{dy}{dt}= y+ Q_2(x,y),
\end{array}
\right.
\end{equation*}
where $P_2$ and $Q_2$ are functions satisfying $P_2(0,0)=Q_2(0,0)=P_2,_x(0,0)=Q_2,_x(0,0)=P_2,_y(0,0)=Q_2,_y(0,0)=0$.
By the implicit function theorem, the equation $y+ Q_2(x,y)=0$ has a solution $y=\varphi(x)$ in a neighborhood of $O$. If we define the function $\psi(x)=P_2(x,\varphi(x))$ then its truncated series expansion will have the form
\begin{equation}
\psi(x)= \Delta_n x^n+...
\end{equation}
where $n\geq 2$ and $\Delta_n\neq 0$. Then it can be shown (\cite{Andronov}, Theor.65, p.340) that
\begin{itemize}
\item If n is odd and $\Delta_n>0$, $(0,0)$ is a \emph{topological node}, i.e., there is a trivial sectorial decomposition consisting in only one parabolic sector.
\item If n is odd and $\Delta_n<0$, $(0,0)$ is a \emph{topological saddle}, i.e., four hyperbolic sectors separated by four separatrices. Two of these separatrices tend to $(0,0)$ in the directions 0 and $\pi$, the other two in the directions $\pi/2$ and $3\pi/2$.
\item If n is even then (0,0) is a \emph{saddle node}, i.e., one parabolic and two hyperbolic sectors separated by three separatrices.
    \begin{itemize}
    \item If $\Delta_n<0$, the hyperbolic sectors contain a segment of the positive x-axis.
    \item If $\Delta_n>0$, the hyperbolic sectors contain a segment of the negative x-axis.
    \end{itemize}
\end{itemize}

\section{Nilpotent critical points}
Let us assume that the origin is an isolated critical point of a planar system and that $A$ (its linearization matrix) is not the zero matrix, but $\det A=0$ and its trace satisfies $T=0$ then the critical point will be \textit{nilpotent}\cite{DLLA}. Again we refer the reader to \cite{Andronov} \cite{DLLA} for details on the theory of the qualitative behaviour of these points. We will summarize here only the main results:

Given a system possessing a nilpotent critical point there is a suitable linear transformation that allows us to write it as:
\begin{equation*}
\left\{
\begin{array}{l}
 \frac{dx}{dt}= y+ P_2(x,y)\\
\frac{dy}{dt}= Q_2(x,y),
\end{array}
\right.
\end{equation*}
where $P_2$ and $Q_2$ are functions satisfying $P_2(0,0)=Q_2(0,0)=P_2,_x(0,0)=Q_2,_x(0,0)=P_2,_y(0,0)=Q_2,_y(0,0)=0$.
By the implicit function theorem, the equation $y+ P_2(x,y)=0$ has a solution $y=\varphi(x)$ in a neighborhood of $O$. If we define the function $\psi(x)=Q_2(x,\varphi(x))$ then its truncated series expansion will have the form
\begin{equation}
\psi(x)= a_k x^k+...
\end{equation}
where $k\geq 2$ and $a_k\neq 0$. On the other hand, if we define $\sigma(x)=P_2,_x(x,\varphi(x))+Q_2,_y(x,\varphi(x))$ then either its truncated series expansion can be written as
\begin{equation}
\sigma(x)=b_n x^n+...
\end{equation}
where $b_n\neq 0$ or $\sigma(x)\equiv 0$ in which case $b_n=0\ \forall n$. Then it can be shown that
\begin{itemize}
\item  If $k$ is \textbf{odd} (\cite{Andronov}, Theor.66, p.357) then we define $p\equiv(k-1)/2$ and $\lambda\equiv b_n^2+2(k+1) a_k$ \begin{itemize}
    \item If $a_k>0$ then $O$ is a \emph{topological saddle} point.
    \item If $a_k<0$ then $O$ is a
        \begin{itemize}
        \item \emph{Focus or center} if either 1) $b_n=0$, 2) $b_n\neq0$ and $n>p$ or 3) $b_n\neq0$, $n=p$ and $\lambda<0$.
        \item \emph{Topological node} if either 1) $n$ is even, $b_n\neq0$ and $n<p$ or 2) $n$ is even, $b_n\neq0$, $n=p$ and $\lambda\geq0$.
        \item Equilibrium state with an \emph{elliptic region}, i.e., an elliptic sector and a hyperbolic sector separated by two separatrices, if either 1) $n$ is odd, $b_n\neq0$ and $n<p$ or 2) $n$ is odd, $b_n\neq0$ and $n=p$ and $\lambda\geq0$.
        \end{itemize}
    \end{itemize}
\item If $k$ is \textbf{even} (\cite{Andronov}, Theor.67, p.362) then we define $p\equiv k/2$ and the critical point is a
    \begin{itemize}
    \item \emph{Cusp} (i.e., two hyperbolic sectors separated by two separatrices) if either $b_n=0$ or $b_n\neq0$ and $n\geq p$
    \item \emph{Saddle node} if $b_n\neq0$ and $n< p$
    \end{itemize}
\end{itemize}

\end{document}